\pdfoutput=1
\documentclass{appolb}

\usepackage{hyperref}
\usepackage{epsfig}
\usepackage{amsmath}
\usepackage{amssymb}
\usepackage{subfigure}

\usepackage{amsbsy}
\usepackage{amsmath}
\usepackage{amssymb}
\usepackage{euscript}
\usepackage{amsmath}

\newcommand{\Peu}{\EuScript{P}}

\newcommand{\Pbbm}{\mathbf{P}}

\newcommand{\veps}{\varepsilon}

\newcommand{\bk}{{\bf{k}}}

\begin{document}

\title{
\vspace{-20mm}
\begin{flushright} \bf IFJPAN-IV-2009-3\\ \end{flushright}
\vspace{5mm}
\uppercase{NLO} \uppercase{QCD} Evolution in the Fully Unintegrated Form%
\thanks{
  This work is partly supported by the EU 
  Framework Programme grants MRTN-CT-2006-035505 and
  MTKD-CT-2004-014319
  and by the Polish Ministry of Science and Higher Education grant 
  No.\ 153/6.PR UE/2007/7.
  \hfill \\
  Presented at the
  {\em Cracow Epiphany Conference
  on Hadronic Interactions at the Dawn of the LHC},
  January 5-7, 2009.}%
}%
\author{S. Jadach and M. Skrzypek
\address{Institute of Nuclear Physics, Polish Academy of Sciences,\\
ul. Radzikowskiego 152, 31-342, Krak\'ow, Poland}}
\maketitle

\begin{abstract}
{\em Abstract:}
The next-to-leading order (NLO) evolution 
of the parton distribution functions (PDF's) in QCD is the ``industry standard''
in the lepton-hadron and hadron-hadron collider data analysis.
The standard NLO DGLAP evolution is formulated for inclusive (integrated)
PDFs and is done using inclusive NLO kernels.
We report here on the ongoing project, called KRKMC,
in which NLO DGLAP evolution is performed for 
the exclusive multiparton (fully unintegrated) distributions (ePDF's)
with the help of the exclusive kernels.
These kernels are calculated within the two-parton phase space
for bremsstrahlung subset of the Feynman diagrams of the non-singlet evolution,
using Curci-Furmanski-Petronzio factorization scheme.
The multiparton distribution with multiple use of
the exclusive NLO kernels is implemented in the Monte Carlo program
simulating multi-gluon emission from single quark emitter.
With high statistics tests ($\sim 10^{9}$ events) it is shown that
the new scheme works perfectly well in practice and is
equivalent at the inclusive level with the traditional inclusive
NLO DGLAP evolution.
Once completed, this Monte Carlo module is aimed as a building
block for the NLO parton shower Monte Carlo, for $W/Z$ production
at LHC and for $ep$ scattering, as well as a starting point
for other perturbative QCD based Monte Carlo projects.

\vspace{3mm}
\centerline{\em Submitted To Acta Physica Polonica B}
\end{abstract}


\vspace{5mm}
\begin{flushleft}
\bf IFJPAN-IV-2009-3\\
\end{flushleft}

\newpage
\section{Introduction}

The so called next-to-leading-order (NLO) parton shower Monte Carlo (MC),
usually referred to as a highly desirable type of the calculation tool
for perturbative QCD predictions at LHC
is commonly believed to be unfeasible in practice.
The ongoing project presented here demonstrates proof of the existence of
such a NLO parton shower MC for the initial state QCD,
albeit for a limited subset of diagrams
of the non-singlet NLO QCD evolution
of the parton distributions functions (PDFs)
The project involves re-calculation of the NLO evolution kernels
in the exclusive (fully unintegrated) form,
following Curci-Furmanski-Petronzio method~\cite{Curci:1980uw}
for the DGLAP type~\cite{DGLAP} of the QCD evolution of PDFs.
A prototype MC with the new exclusive kernels
performs {\em exactly} the NLO DGLAP evolution of PDFs on its own
(no external pretabulated PDFs needed!),
following closely the collinear factorization theorems
of QCD and the standard ``Matrix Element $\times$ Phase Space'' approach.
Once completed, this project
will open new avenues for a new class of perturbative QCD calculations
in form of Monte Carlo event generators
for the coming two decades of the LHC experiments.

What are the main aims and assumptions of the project?
The new NLO Parton Shower Monte Carlo for QCD
initial state radiation should be:
\begin{itemize}
\item
based firmly on Feynman diagrams (matrix element) defined within
the standard Lorentz invariant phase space,
\item
based rigorously on the collinear factorization of the classic
works, see~\cite{Ellis:1978sf,Curci:1980uw} or~\cite{Collins:1984kg},
\item
implementing exactly NLO DGLAP evolution in $\overline{MS}$ scheme,
\item
implementing the exclusive PDF (ePDF) defined within the standard
multiparton Lorentz invariant phase space%
\footnote{Similar kind of PDF is also referred to as
  a ``fully unintegrated'' PDF~\cite{Collins:2007ph}.},
\item
performing NLO evolution by the MC itself,
with help of the new exclusive NLO kernels.
\end{itemize}
The above priorities can be compared with ref.~\cite{Collins:2000gd},
where construction of NLO parton shower MC is also advocated.

In the realization of the presented KRKMC project
very strong emphasis is put from the very beginning
on testing all theoretical and practical ideas with
series of high quality numerical tests.
In fact, every major milestone of the project realization
is marked by the construction and testing the corresponding numerical
MC prototype.
In this contribution we shall report results form
first few MC exercises of this kind.
Unfortunately, we shall not be able to give all the details
of these MC constructions in this short presentation --
we hope to report them in detail in separate publications soon.

Let us attempt to describe very briefly the basic features of the theoretical
QCD model of the parton distributions in the KRKMC NLO MC.
In the folowing formula implemented in the KRKMC program,
\begin{equation}
\label{eq:ePDFdef}
  D(x,Q)
    =\sum_{n=0}^\infty \int \prod_{i=1}^n \frac{d^3 k_i}{2k^0_i}\;
    \tilde{\rho}^{(n)}(P;k_1,k_2\dots,k_n)\;
    \Theta_{S(k_1,k_2\dots,k_n)<Q}\;
    \delta_{1-x=\sum \alpha_i},
\end{equation}
the exclusive parton distribution function, ePDF,
is the integrand $\tilde{\rho}^{(n)}$.
In the above we denote the lightcone variables of the emitted partons
as $\alpha_i=\frac{k_i^+}{2E}=\frac{k_i^0+k_i^3}{2E}$,
where $k^\mu_i=(k_i^0,\bk_i,k_i^3),\; i=1,..,n$ are 4-momenta of these partons,
while the initial parton 4-momentum is $q_0=P=(E,0,0,E)$.
For the variable $S$ (factorization scale)
we use one of the following variables
\begin{equation}
\label{eq:Sdef}
\begin{split}
&S_q=\bigg|\bigg(P-\sum\limits_{i=1}^n k_i\bigg)^2\bigg|^{1/2},
\qquad
S_r=\max\left(\frac{|\bk_1|}{\alpha_1},\frac{|\bk_2|}{\alpha_2},
           \dots \frac{|\bk_n|}{\alpha_n}\right),
\\&
S_v=\max\left(\frac{|\bk_1|}{\sqrt{\alpha_1}},\frac{|\bk_2|}{\sqrt{\alpha_2}},
           \dots \frac{|\bk_n|}{\sqrt{\alpha_n}}\right),
\qquad
S_T=\max(|\bk_1|,|\bk_2|,\dots, |\bk_n|),
\end{split}
\end{equation}
$S_q=Q$ is used in ref.~\cite{Curci:1980uw}, while
in the following we shall usually work with maximum transverse momentum $S_T$.
Variable $S_r$ related to maximum rapidity of the emitted real partons
is generally our preferred choice
and occasionally we shall use maximum lightcone minus variable $S_v$.

The integrands in eq.~(\ref{eq:ePDFdef}) are finite due to
built-in regulators of the collinear divergences.
The actual integrands $\tilde\rho^{(n)}$ in eq.~(\ref{eq:ePDFdef})
are the result of projecting 
to the LO+NLO level
the differential distribution
\begin{equation}
\label{eq:projection}
   \rho^{(n)}(P;k_1,k_2\dots,k_n) \to \tilde\rho^{(n)}(P;k_1,k_2\dots,k_n),
\end{equation}
where $\rho^{(n)}$ is coming from the UV subtracted Feynman diagrams%
\footnote{ It includes IR regulators of the collinear and soft singularities.}
using methodology of the ``factorization theorems'' 
of QCD in the physical gauge,
such that the inclusive $D(x,Q)$ obeys {\em exactly}
the standard evolution equation
\begin{equation}
\label{eq:evoleq}
  \frac{\partial}{d\ln Q} D(x,Q)
        = \int dz du\; \Peu^{LO+NLO}(Q,z)\; D(u,Q)\; \delta_{x=zu}
\end{equation}
of the NLO DGLAP with the standard inclusive kernels
of the $\overline{MS}$ scheme
\begin{equation}
\Peu^{LO+NLO}(Q,z)
    = \frac{\alpha(Q)}{2\pi} P^{(0)}(z)
    + \left(\frac{\alpha(Q)}{2\pi}\right)^2 P^{(1)}(z).
\end{equation}
How to perform the projection of eq.~(\ref{eq:projection})
is of course the main theoretical issue in the project.
In particular even the existence of this projection,
such that eq.~(\ref{eq:evoleq}) is fulfilled,
is highly nontrivial and still open question.

On one hand, we insist very strongly on the perfect compatibility of our ePDF's
with the NLO DGLAP in $\overline{MS}$ scheme in the sense of eq.~(\ref{eq:evoleq})
and we believe that it is feasible.
On the other hand, this requirement should not be overstressed,
as we do foresee in the next step going beyond NLO DGLAP in
the exclusive form, for instance towards BFKL.
However, in our opinion such an extension 
has to wait until the exclusive NLO DGLAP evolution
in the MC form is well established and tested.

The multiparton distribution $\tilde\rho^{(n)}$ describes
the chain of the parton emissions out of the {\em single emitter} parton
coming from the hadron beam and absorbed in the hard process.
In the classic LO parton shower MC $\tilde\rho^{(n)}$ is rather simple,
just the product of the LO kernels times Sudakov formfactor.
Momenta $k_i^\mu$ of the MC event are generated according
to LO level $\tilde\rho^{(n)}$, typically using backward evolution
algorithm of ref.~\cite{Sjostrand:1985xi},
or the new {\em constrained MC algorithms}
of refs.~\cite{Jadach:2005bf,Jadach:2005yq}.
On the other hand, the new and more complicated
$\tilde\rho^{(n)}$ at the NLO level
will be basically the convolution of the 2PI kernels
of the collinear factorization scheme
of refs.~\cite{Ellis:1978sf,Curci:1980uw},
truncated at the complete NLO level (in the axial gauge).
Of course, $\tilde\rho^{(n)}$ has to be defined in four dimensions,
$d=4$, and the removal the IR dimensional regulator from the real emission
part of the Curci-Furmanski-Petronzio scheme~\cite{Curci:1980uw},
without spoiling main features of the standard $\overline{MS}$ scheme,
is one of the main practical issues.

Since the main goal of
the KRKMC project is to reproduce NLO DGLAP evolution in QCD
in the exclusive way, in the Monte Carlo,
while its standard inclusive version is a well established technique
and serves well to describe lepton-hadron and hadron-hadron collisions,
it is therefore a valid question to ask:
``why bother to do the same in more complicated way''.
The answer is that it is worth to redo it in the exclusive MC form,
because there are numerous potential gains once it is available;
it may serve either as a building block in a bigger new MC project,
or as an excellent starting point for many new types of the MC
parton shower type MC projects.
Let us list some of the possible applications:
\begin{itemize}
\item
Extension towards CCFM~\cite{CCFM}, BFKL~\cite{BFKL}
(low $x$ limit in PDFs),
with the correct soft limit and built-in colour coherence.
\item
More realistic description of
the heavy quark thresholds in PDFs and the hard process.
\item
The use of exact amplitudes for multiparton emission in the soft
and collinear limits --
the analog of Coherent Exclusive Exponentiation in
QED~\cite{ceex1:1999,Jadach:2000ir}.
\item
Better connection between hard process ME and the shower parts,
as compared with  MC$@$NLO~\cite{Frixione:2002ik}
and the other similar approaches of this class.
In particular we expect no negative weight events,
no ambiguity of defining last emission before hard process, etc.
\item
Providing better tool for exploiting HERA DATA for LHC,
in particular for fitting $F_2$ directly with the MC programs.
\end{itemize}

\begin{figure}[!ht]
\centering
\includegraphics[width=110mm]{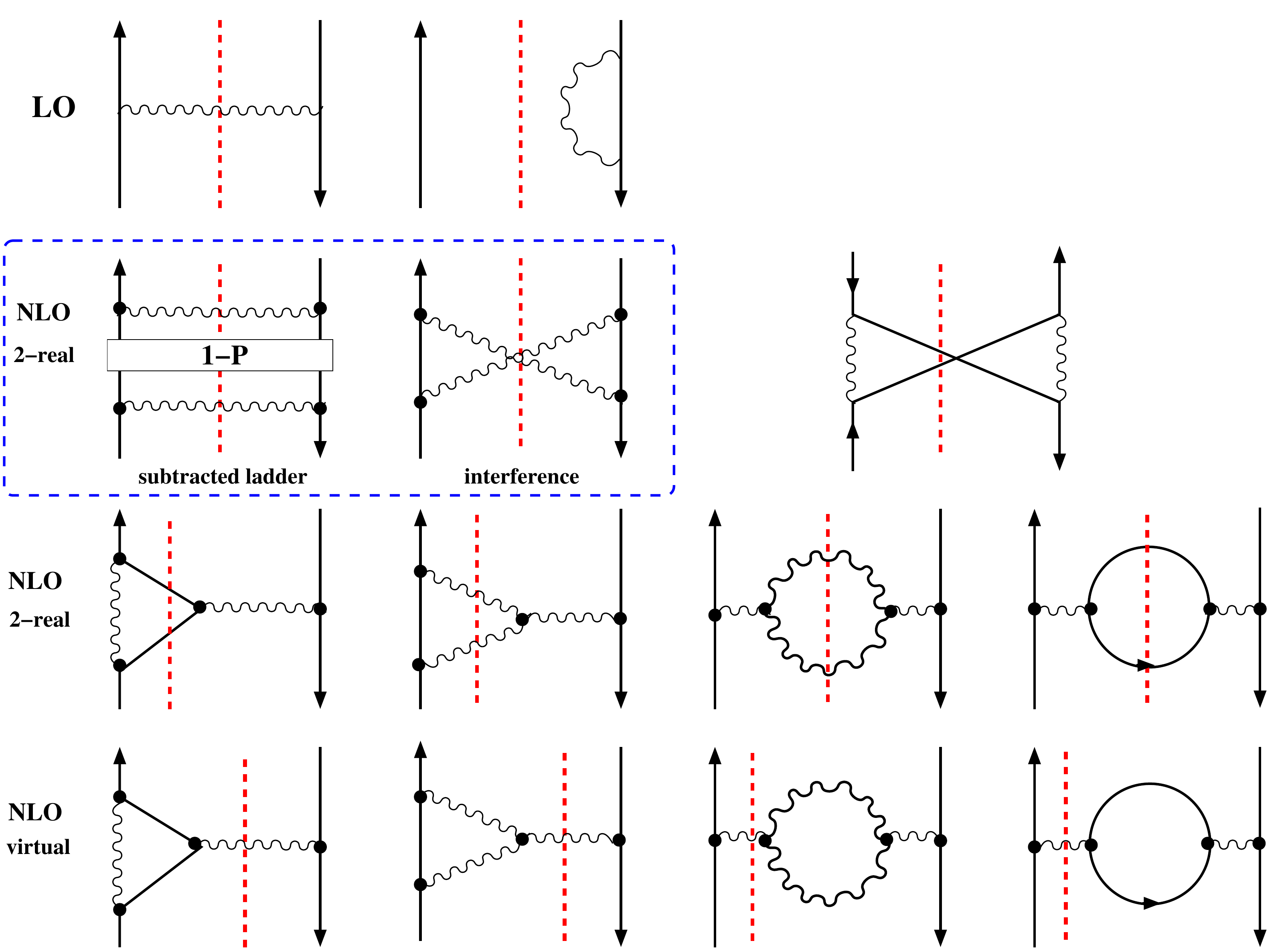}
\caption{
   Feynman diagrams contributing to the calculation of the non-singlet NLO kernel,
   except two-loop self-energy graphs.
}
\label{fig:one}
\end{figure}

\section{Constructing and testing Exclusive kernels}

The first inevitable step in the construction of the multiparton
MC implementing complete NLO DGLAP in the exclusive
way is to construct and test the exclusive NLO evolution kernels
within the two parton phase space in $d=4$, which later on will
be used many times in the $n$-parton emission ladder.
Since our ambition is to be fully compatible with the NLO DGLAP, it is
therefore natural that we take as a starting point the calculation
of the NLO kernels in the diagrammatic way according to
Curci-Furmanski-Petronzio (CFP) scheme~\cite{Curci:1980uw},
rather than technique of calculating anomalous dimensions
of the Wilson operators~\cite{Floratos:1977au,Floratos:1978ny}.
In the CFP scheme one exploits the statement of
ref.~\cite{Ellis:1978sf} (EGMPR)
that in the physical (axial) gauge all collinear singularities are located
on the rungs of the ladder diagram between
the 2PI kernels.
Feynman diagrams contributing to non-singlet NLO kernels
are depicted in Fig.~\ref{fig:one}.
In the following discussion we will limit ourselves
to two bremsstrahlung diagrams
inside the dashed box marked in this figure,
in fact to $C_F^2$ part of them.

\subsection{Re-Calculating NLO DGLAP kernels following CFP}
Recalculation of the NLO DGLAP kernels is mandatory because
we need access to the two parton differential distributions
in the internal phase space of the kernels.
Moreover, we shall need explicit expressions for the virtual contributions
to NLO DGLAP kernels from virtual Feynman diagrams, which are not
available in ref.~\cite{Curci:1980uw},
nor in the papers \cite{Ellis:1996nn,Bassetto:1998uv},
where the calculations of refs.~\cite{Curci:1980uw,Furmanski:1980cm}
were reproduced and cross-checked once again.

Collinear factorization theorem of EGMPR~\cite{Ellis:1978sf}
improved and customized to the use of $\overline{MS}$
dimensional regularization scheme
by CFP~\cite{Curci:1980uw} states that
\begin{equation}
\begin{split}
&F=C_0\cdot  \frac{1}{1-K_0}
=C\left(\alpha,\frac{Q^2}{\mu^2} \right)
\otimes \Gamma\left(\alpha, \frac{1}{\epsilon}\right),
\\&~~~~~~~
 = \left\{ C_0\cdot \frac{1}{1-(1-\Pbbm)\cdot K_0} \right\}\otimes
\left\{
  \frac{1}{1-\left(\Pbbm K_0\cdot\frac{1}{1-(1-\Pbbm)\cdot K_0} \right )}
\right\}_\otimes,
\\&
\Gamma\left(\alpha, \frac{1}{\epsilon}\right)
 \equiv \left(\frac{1}{1- K}\right)_\otimes
 = 1+K+K\otimes K +K\otimes K\otimes K+...,
\\&
K=\Pbbm K_0\cdot\frac{1}{1-(1-\Pbbm)\cdot K_0},
\quad
C=C_0\cdot \frac{1}{1-(1-\Pbbm)\cdot K_0}.
\end{split}
\end{equation}
where $K_0$ is 2-particle irreducible (2PI) kernel,
which is free of the collinear divergences.
The ladder part $\Gamma$ corresponds to the MC parton shower
and $C$ is the hard process part.
The projection operator of CFP scheme is defined as follows
\begin{equation}
 \Pbbm = P_{spin}\; P_{kin}\; PP
\end{equation}
and it consists of the kinematic projection operator $P_{kin}$,
spin projection (averaging) operator $P_{spin}$
and the pole part $PP$ extracting $\frac{1}{\epsilon^k}$, $k>0$ part%
\footnote{The above formula should be treated with special care,
   see ref.~\cite{Curci:1980uw}, due to non-associativity of the products
   of the kernels with insertions of the $(1-\Pbbm)$ operator.}.
Multiplication symbol $\cdot$ denotes
full phase space integration $d^n k$ (the only source of collinear divergences),
while convolution $\otimes$ involves the integration
over the 1-dimensional lightcone variable $x$ only.

In the CFP methodology
the NLO kernel is extracted from the second order term in the expansion:
\begin{equation}
\begin{split}
&\Gamma =\frac{1}{1- K}
= 1+\Pbbm K_0 +\Pbbm K_0 \cdot [(1-\Pbbm)\cdot K_0]
  +[\Pbbm K_0]\otimes[\Pbbm K_0]+\cdots,
\\&
C =C_0 \frac{1}{1-(1-\Pbbm) K_0}
\\&~~~
=
C_0+ C_0\cdot (1-\Pbbm)\cdot K_0 
   +C_0\cdot (1-\Pbbm)\cdot K_0 \cdot [(1-\Pbbm)\cdot K_0)]+\dots
\end{split}
\end{equation}
Following the ``pole-part method'' of CFP
the NLO kernel is obtained as follows%
\footnote{$Res_1$ denotes coefficient in front of $\frac{1}{\veps}$ pole
   in the Laurent expansion around $\veps=0$.}:
\begin{equation}
\begin{split}
& \Peu(x)=
 \frac{\alpha}{2\pi} P(\alpha,x)
= \frac{\alpha}{2\pi} P^{(0)}(x) 
 +\left(\frac{\alpha}{2\pi}\right)^2 P^{(1)}(x)
= 
  \alpha\; \frac{\partial }{\partial \alpha} 
  Res_1\; \Gamma(\alpha,x)
\\&
= Res_1\Big\{ \Pbbm K_0 \Big\}
 +2 Res_1\Big\{ \Pbbm K_0 \cdot [(1-\Pbbm)\cdot K_0] \Big\}.
\end{split}
\end{equation}

\begin{figure}[!ht]
\centering
\includegraphics[width=110mm]{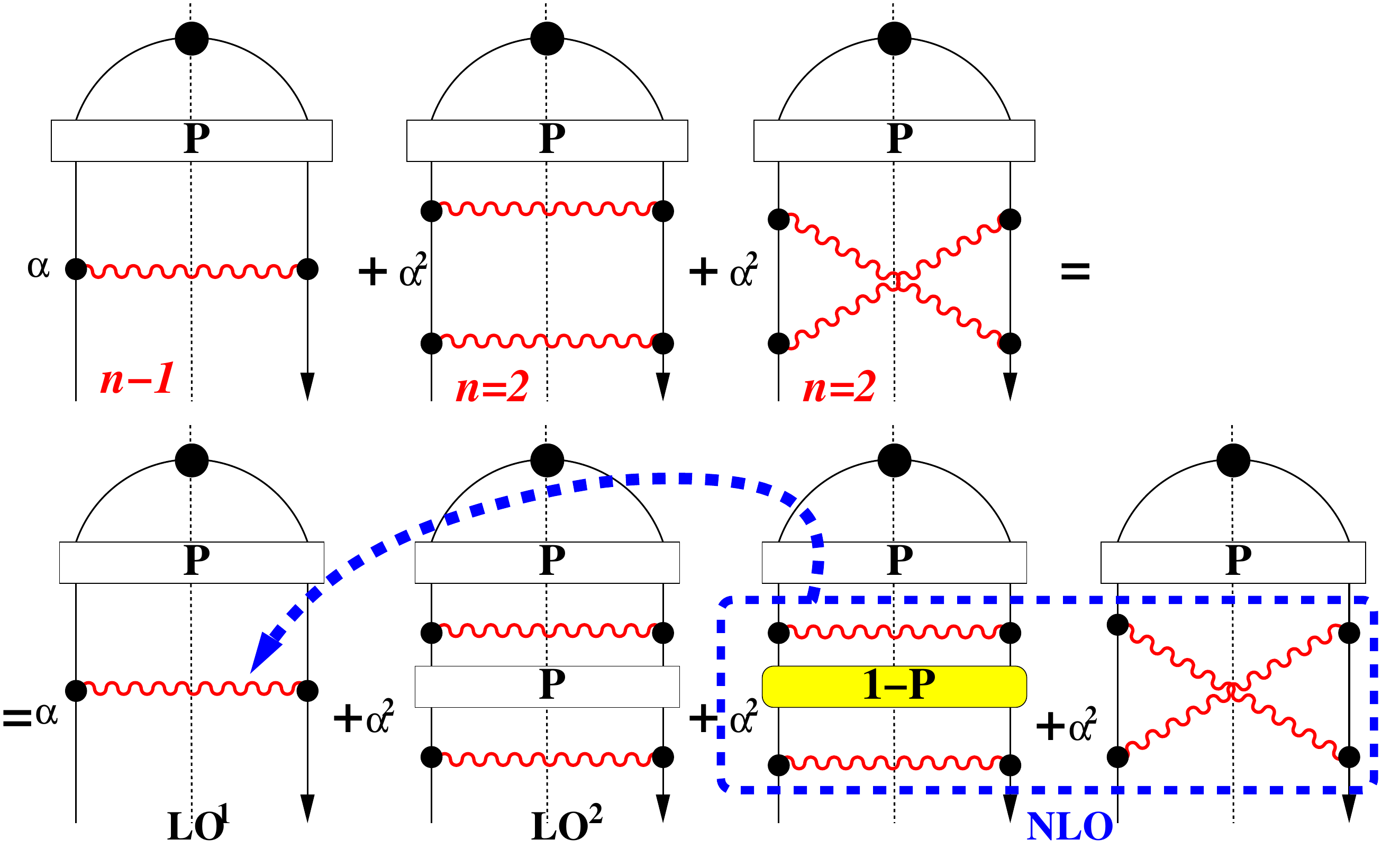}
\caption{
   Extraction of NLO kernel for bremsstrahlung diagrams.
}
\label{fig:two}
\end{figure}

The role of the $\Pbbm$ projection operator in the CFP scheme in the case of
our bremsstrahlung diagrams is illustrated in Fig.~\ref{fig:two}.
The upper line in Fig.~\ref{fig:two} shows the three Feynman
diagrams, defined in the standard phase space which is closed from above
by the operator $\Pbbm$.
It employs also cut-off parameter $Q=\mu_F$ to limit phase space from the above.
The well known property of the physical gauge~\cite{lipatov:75}
tells us that 2nd diagram in Fig.~\ref{fig:two} contributes
double collinear log (or $1/\veps^2$), while the 1st
and 3rd only single collinear log (or $1/\veps$).
The insertion of $\Pbbm+(1-\Pbbm)$ in between two gluon rungs in the 2nd
diagram changes nothing in the total sum, however, the part containing
$(1-\Pbbm)$ is now only single logarithmic, and in the factorization procedure
it is combined with the last ``crossed diagram'', 
forming together the NLO contribution to evolution kernel,
as depicted in the lower part of Fig.~\ref{fig:two}.
Let us remark that the remaining part of 2nd diagram with two gluon rungs
and two $\Pbbm$'s is just the square of the 1st diagram,
so it represents the ``idealized'' 2nd order LO contribution
(the beginning of the LO series).
The NLO part (dashed box)
is integrated over the 2-gluon phase space%
\footnote{Constraining total loss of the lightcone variable.}
and associated with the single gluon 1st diagram (see dashed arrow),
forming the NLO part of the standard inclusive DGLAP kernel.
We want to stress that this procedure combines 1 and 2 real gluon contributions
(plus virtual ones) into a single inclusive object,
the standard NLO inclusive kernel.

Our main aim
is to recover the internal 2-gluon differential distribution
of the NLO kernel in the MC event generator.
{\em In a sense our aim is to undo the procedure marked 
by the dashed arrow in Fig.~\ref{fig:two}.}
Let us now concentrate on the above internal 2-gluon differential distribution
of the NLO kernel for the bremsstrahlung diagrams
($C_F^2$ part).

\subsection{Two-gluon differential distribution inside NLO DGLAP kernel}
\label{sec:two}
\begin{figure}[!ht]
\centering
\includegraphics[width=90mm]{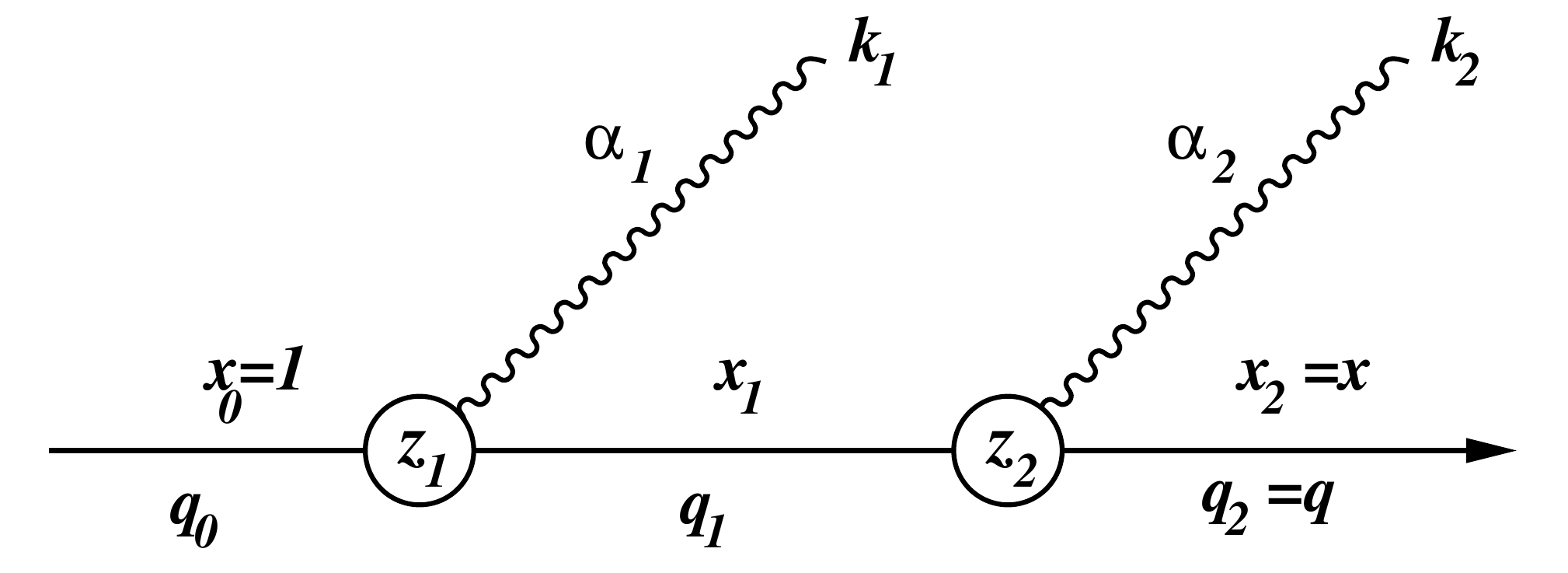}
\caption{
   Kinematics of the 2 gluon emission.
}
\label{fig:kinema}
\end{figure}
Before we describe the 2-gluon differential distribution let us define kinematics
of the process, see also Fig.~\ref{fig:kinema}:
\begin{equation}
\begin{split}
& x_i     = \frac{\zeta\cdot q_i}{\zeta\cdot q_0},\quad
\alpha_i= \frac{\zeta\cdot k_i}{\zeta\cdot q_0}=\frac{k^+}{2E} ,\quad
q_i= q_0-\sum_{j=1}^i k_j,
\\&
q_0^2=0,\quad P=q_0=(E,0,0,E),
\\&
\zeta=(1,0,0,-1),\quad \zeta^2=0,
\\&
k_i^2=0,\quad k_i=(k_i^0,\bk_i,k_i^3).
\end{split}
\end{equation}

The second order contribution to $\Gamma$,
for the 2nd bremsstrahlung diagram in Fig.~\ref{fig:two},
in $d=4+2\veps$ dimensions, reads
\begin{equation}
\label{eq:Cbar}
\begin{split}
&\bar{C}_0K_0K_0 
=C_F^2 \alpha^2
\int \frac{d\alpha_1}{\alpha_1}
     \frac{d\alpha_2}{\alpha_2}
     \delta_{1-x=\alpha_1+\alpha_2}\;
\int d^{2+2\epsilon}\bk_1 d^{2+2\epsilon}\bk_2\;
     \mu^{-4\epsilon}\;
     \Theta(S(k_1,k_2)<Q)
\\&~~~\times
\left\{  
         \frac{T_1(\alpha_1,\alpha_2)}{\alpha_1\alpha_2}
        +\frac{T_2(\alpha_1,\alpha_2,\epsilon)}{\alpha_2^2}
               \frac{\bk_2^2}{\bk_1^2}\;
        +\frac{T_3(\alpha_1,\alpha_2)}{\alpha_2}
               \frac{2\bk_1 \cdot\bk_2}{\bk_1^2}\;
\right\}
\frac{1}{q^4(k_1,k_2)},
\end{split}
\end{equation}
with the notation explained below
and the extracted kernel being independent
of the actual choice of $\bar{C}_0$ in eq.~(\ref{eq:Cbar})%
\footnote{We use $\bar{C}_0=P_{spin}\theta_{S(k_1,k_2,...,k_n)<Q}$,
  where $S$ is one of these in eqs.~(\ref{eq:Sdef}).}.
The $C_F^2$ part of the NLO kernel
extracted from $\Gamma$ function
for all 3 bremsstrahlung diagrams including 2 soft counterterms
reads
\begin{equation}
\label{eq:PN}
\Peu^N(x)
=\frac{1}{2!}
 \int \frac{d^3 k_2}{2 k^0_2}
 \int \frac{d^3 k_1}{2 k^0_1}\;
   \delta_{1=\max(|\bk_1|,|\bk_2|)/Q}\;
   \delta_{1-x=\alpha_1+\alpha_2}\;
   b^N_2(k_1,k_2),
\end{equation}
where
\begin{equation}
\label{eq:b2N}
\begin{split}
&b^N_2(k_1,k_2)= 
\frac{(\alpha C_F)^2}{16(2\pi)^2} 
\Big[
 b^{Ladd.}(k_1,k_2)
-b^{Count.}(k_1,k_2)
\\&~~~~~~~~~~~~~~~~~~~~~~~~~
+b^{Ladd.}(k_2,k_1)
-b^{Count.}(k_2,k_1)
+b^{XLad.}(k_1,k_2)
\Big],
\\&
b^{Ladd.}(k_1,k_2)=
\frac{1}{q^4(k_1,k_2)}
\left\{  
         \frac{T_1(\alpha_1,\alpha_2)}{\alpha_1\alpha_2}
        +\frac{T_2(\alpha_1,\alpha_2,0)}{\alpha_2^2}
               \frac{\bk_2^2}{\bk_1^2}
        +\frac{T_3(\alpha_1,\alpha_2)}{\alpha_2}
               \frac{2\bk_1 \cdot\bk_2}{\bk_1^2}
\right\},
\\&
b^{Xlad.}(k_1,k_2)=\frac{1}{q^4(k_1,k_2)}
\Bigg\{ \frac{2T^x_1(\alpha_1,\alpha_2)}{\alpha_1\alpha_2}
\\&~~~~~~~
   +T^x_{2}(\alpha_1,\alpha_2) \frac{2\bk_1\cdot\bk_2}{\alpha_1\bk_2^2}
   +T^x_{2}(\alpha_2,\alpha_1) \frac{2\bk_1\cdot\bk_2}{\alpha_2\bk_1^2}
   +T^x_3(\alpha_1,\alpha_2)
            \frac{(\bk_1 \cdot\bk_2)^2}{\bk_1^2\bk_2^2}
\Bigg\},
\\&
b^{Count.}(k_1,k_2)=
   \frac{\bk_2^2}{q^4(0,k_2)}
   \frac{T_2(\alpha_1,\alpha_2,0)}{\alpha_2^2}
   \frac{1}{\bk_1^2}
   \theta_{\bk_1^2<\bk_2^2}
=  \frac{T_2(\alpha_1,\alpha_2,0)}{(1-\alpha_1)^2 \bk_1^2 \bk_2^2}
   \theta_{\bk_1^2<\bk_2^2},
\\&
   -q^2(k_1,k_2)=
      \frac{1-\alpha_2}{\alpha_1} \bk_1^2
     +\frac{1-\alpha_1}{\alpha_2} \bk_2^2
     +2\bk_1 \cdot \bk_2.
\end{split}
\end{equation}
where $\gamma$-trace factors and the second quark propagator are
\begin{equation}
\begin{split}
&  T_1(\alpha_1,\alpha_2) =(1+x^2+x_1^2)\alpha_1\alpha_2,\quad
   T_3(\alpha_1,\alpha_2)  = (1+x^2+x_1^2)x_1,
\\&
   T_2(\alpha_1,\alpha_2,\epsilon) = (1+x_1^2)(x^2+x_1^2)
        +\epsilon T_2'(\alpha_1,\alpha_2),
\\&
   T_2'(\alpha_1,\alpha_2)
     =(1-x_1)^2(x^2+x_1^2) +(x-x_1)^2(1+x_1^2),
\\&
   T^x_1(\alpha_1,\alpha_2)   =x(1+x^2 -\alpha_1\alpha_2),\qquad
   T^x_3(\alpha_1,\alpha_2) =2(1+x^2),
\\&
  T^x_{2}(\alpha_1,\alpha_2)= x(1-\alpha_1) +(1+x^2)(1-\alpha_2),
\end{split}
\end{equation}
The integral (\ref{eq:b2N}) is already in $d=4$ dimensions and it is finite.
For simplicity we have omitted term%
\footnote{We keep them in the actual program 
   and in the following numerical tests.}
$T'_2$ from the $\epsilon$ part in the $\gamma$-trace leading to $T_2$.

\begin{figure}[!ht]
\centering
   \includegraphics[width=120mm,height=70mm]%
            {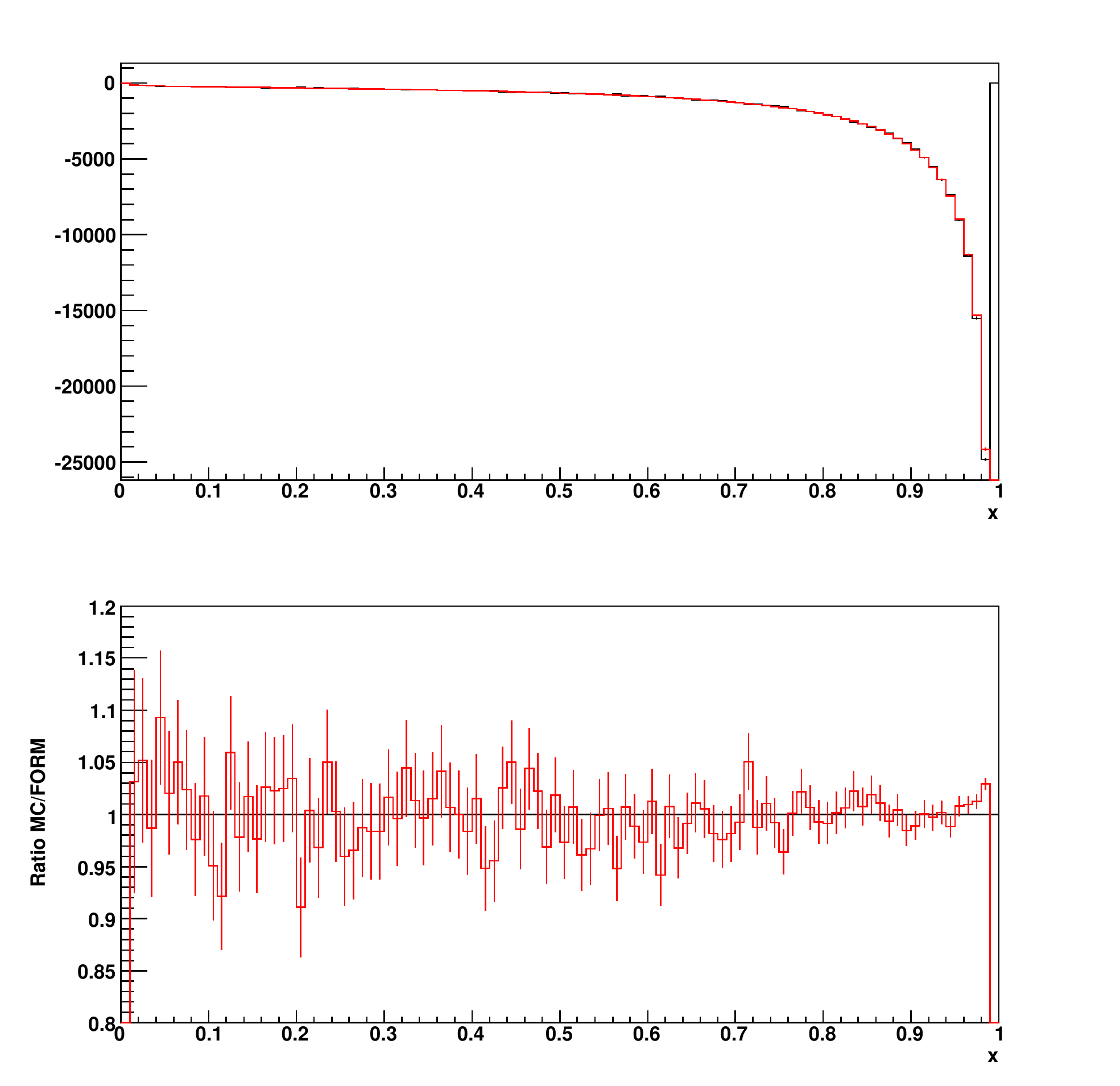}
\caption{
   Comparison of numerical MC and analytical integration of $b^N_2(k_1,k_2)$
   contribution to exclusive NLO kernel.
}
\label{fig:PN}
\end{figure}

The integrations in eqs.~(\ref{eq:PN}) have been performed analytically:
\begin{equation}
\label{eq:PNan}
\Peu^N(z)=
 \left(C_F^2\frac{\alpha}{\pi}\right)^2
 \left(
 \frac{1+3x^2}{16(1-x)}\ln^2(x)
   +\frac{2-x}{4}\ln(x) +\frac{3}{8}(1-x)
 \right),
\end{equation}
reproducing the result of CFP~\cite{Curci:1980uw}.
On the other hand 
the distribution $b^N_2(k_1,k_2)$ 
of eq.~(\ref{eq:b2N}) has been plugged into
the general purpose MC integrator/simulator FOAM and
the result of numerical MC integration was compared with
formula of eq.~(\ref{eq:PNan}),
see Fig.\ref{fig:PN}.
Both NLO results agree to within statistical MC relative error of a couple percent.
The same error with respect to LO+NLO is only $\sim 10^{-3}$.

However, our real interest is not the integral but rather the integrand
itself, as it will be used as a building block in the multiparton
distribution for the MC, hence in the following we shall analyze the shape
of $b^N_2(k_1,k_2)$  and its singularity structure in a detail.

\subsection{Analyzing internal structure of the  NLO DGLAP kernel}

\begin{figure}[!ht]
\centering
\includegraphics[width=125mm,height=80mm]{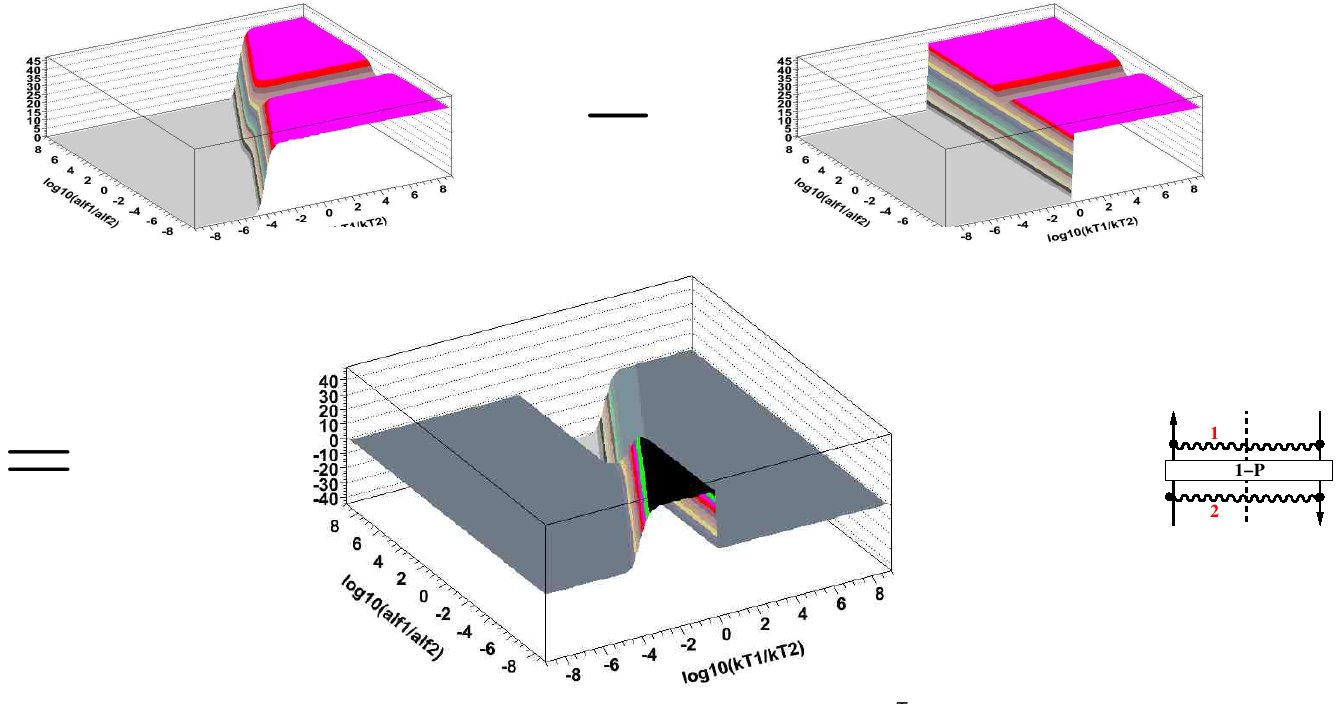}
\caption{
   Contribution to unintegrated NLO kernel from one of the double-bremsstrahlung diagrams
   (top left-hand plot).
   Subtraction of the internal singularity, $k_2^T\to 0$, is done using soft counterterm
   (top right-hand side).
   Subtracted distribution is also shown (bottom).
   Lightcone variable is fixed, $x=0.3$.
}
\label{fig:Br2S}
\end{figure}

The activity of the previous section,
culminated in Fig.\ref{fig:PN} was just the warm-up exercise,
in which we have checked whether we are able to build
a numerical model for the NLO kernel in the exclusive form.
The next step, which we pursue in the following, is to examine
the shape and singularity structure of the integrand.
The final step will be to implement the distribution $b^N_2(k_1,k_2)$
several times in the actual multiparton distribution of the
NLO parton shower MC program, see next Section.
The extension of the present analysis of the IR singularity
structure to more Feynman diagrams and
genuine non-abelian contributions $\sim C_F C_A$
is reported in ref.~\cite{ifjpan-iv-09-2} in these proceedings.

To begin with we check how does the integrand $b^N_2(k_1,k_2)$ look like
in the Sudakov-type variables $\ln(k^T_1/k^T_2)$ and  $\ln(\alpha_1/\alpha_2)$,
keeping two constraints: $\max(k^T_1/k^T_2)=Q$ and $\alpha_1+\alpha_2=1-x$.
In particular we are interested in the soft limit of one of the two gluons,
for instance $k^T_2\to 0$ and $\alpha_2\to0$.
The single diagram of bremsstrahlung with the soft subtraction
\[
b^{Ladd.}(k_2,k_1)-b^{Count.}(k_2,k_1)
\]
is visualized in Fig.~\ref{fig:Br2S}.
In this figure
the upper-left-hand plot represents the distribution
from single Feynman diagram of the double brems\-strahlung type.
The ``cliff'' along the line of equal virtualities of the emitted gluons
$k^T_1/\sqrt{\alpha_1}=k^T_2/\sqrt{\alpha_2}$
represents well know ``ordering effect'' in the physical gauge
due to propagator of the emitter quark.
The soft counterterm of CFP depicted 
in the upper-right-hand plot removes soft singularity
in the limit $k^T_2\to 0$.
However, the remaining structure near $k^T_1 \sim k^T_2$
features very strong cancellations between two triangular regions
of the doubly-logarithmic size.
Such structures, if they were really present, would render any MC implementation
of the NLO unintegrated kernel unfeasible
due to strong MC weight fluctuations.

\begin{figure}[!ht]
\centering
\includegraphics[width=125mm,height=55mm]{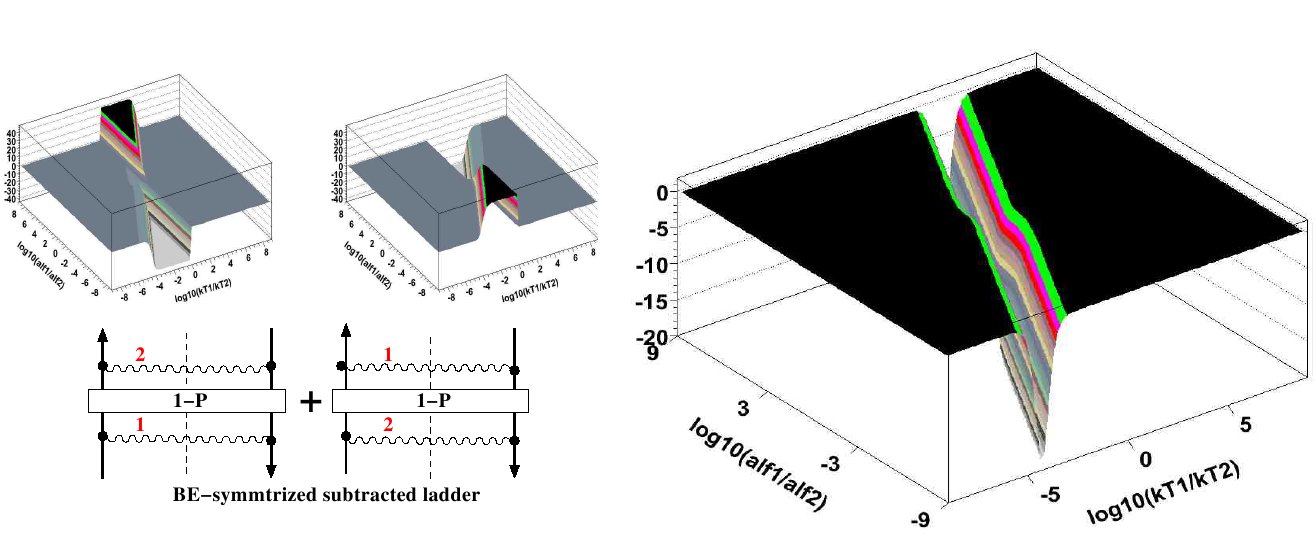}
\caption{
   Contribution to unintegrated NLO kernel from two double-bremsstrahlung diagrams
   with subtraction,  $x=0.3$.
}
\label{fig:Br12S}
\end{figure}

\begin{figure}[!ht]
\centering
\includegraphics[width=125mm,height=85mm]{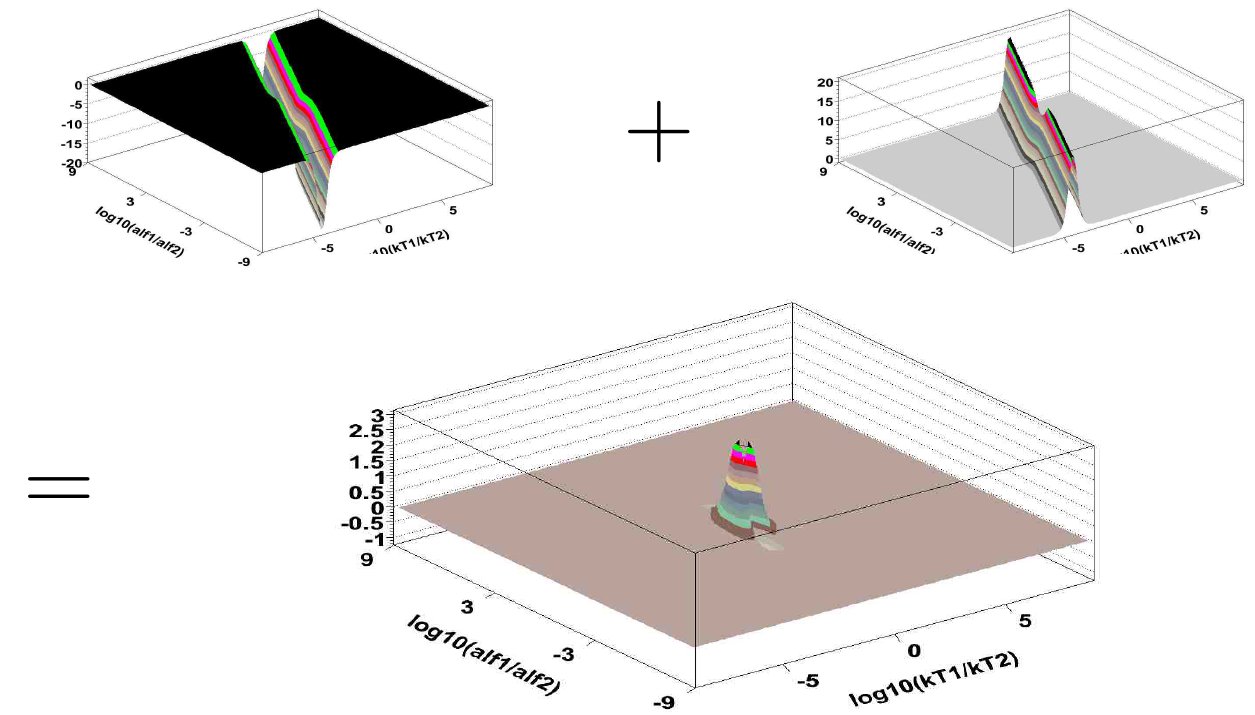}
\caption{
   Contribution to unintegrated NLO kernel 
   from all three double-bremsstrahlung diagrams
   with subtraction. $x=0.3$.
}
\label{fig:BrBxS}
\end{figure}

Fortunately, the inclusion of the second bremsstrahlung diagram
\[
b^{Ladd.}(k_1,k_2)-b^{Count.}(k_1,k_2)   +b^{Ladd.}(k_2,k_1)-b^{Count.}(k_2,k_1)
\]
makes the distribution more friendly for the future MC use,
as shown in Fig.~\ref{fig:Br12S}.
Nevertheless the single-log structure along the line
$k^T_1/\sqrt{\alpha_1}=k^T_2/\sqrt{\alpha_2}$
seems to contradict the eikonalization of the brems\-strahlung distributions
in the soft limit~\cite{yfs:1961}
and would cause the problems for the MC use.
The size of the remaining uncanceled part is much smaller
than the LO distributions.
In this plot we have used 
$x=1-\alpha_1-\alpha_2=0.3$
and the integration over azimuthal angles of the gluons has been performed.
However, we have checked that general features of the above results
remain the same for any values of $x$ and any gluon azimuthal angle.
In ref.~\cite{ifjpan-iv-09-2} similar plot is shown for
$b^{Ladd.}(k_1,k_2)+b^{Ladd.}(k_2,k_1)$,
that is for the bremsstrahlung matrix element
without soft counterterm subtraction.

In Fig.~\ref{fig:BrBxS}.
the above single-logarithmic structure disappears when finally
the ``crossed'' diagram (interference) is included.
As we see, the interference diagram 
(see upper-right-hand plot) features the same
single-log structure along the line of equal virtualities,
which corrects the soft eikonal limit and the sum is nonzero only
in the region of $k^T_1 \sim k^T_2$
and $\alpha_i\sim \frac{1}{2}(1-x) \sim 1$.
In other words, $b^N_2(k_1,k_2)$ plays a role of the ``short range correlation''
function on the Sudakov plane which enters into the game only
when both gluons are not soft and they are not in the so-called
``strong ordering'' regime.
Such a behaviour looks also very friendly for the Monte Carlo use,
if we plan to use $b^N_2(k_1,k_2)$ as a component in the MC weight
correcting LO distribution to the NLO level.

\section{Re-insertion of exclusive NLO kernels into LO Monte Carlo}

Once exclusive representation of the NLO evolution kernel
in two parton phase space is constructed and analyzed,
the next question is how to use
it to construct the multiparton distribution of the PS MC.
We shall show in the following how to solve this problem 
for the $C_F^2$ part of the bremsstrahlung diagrams contributing
to the non-singlet NLO kernel.

\subsection{General scheme of exclusive NLO insertion}
\begin{figure}[!hb]
\centering
   \includegraphics[width=130mm,height=50mm]{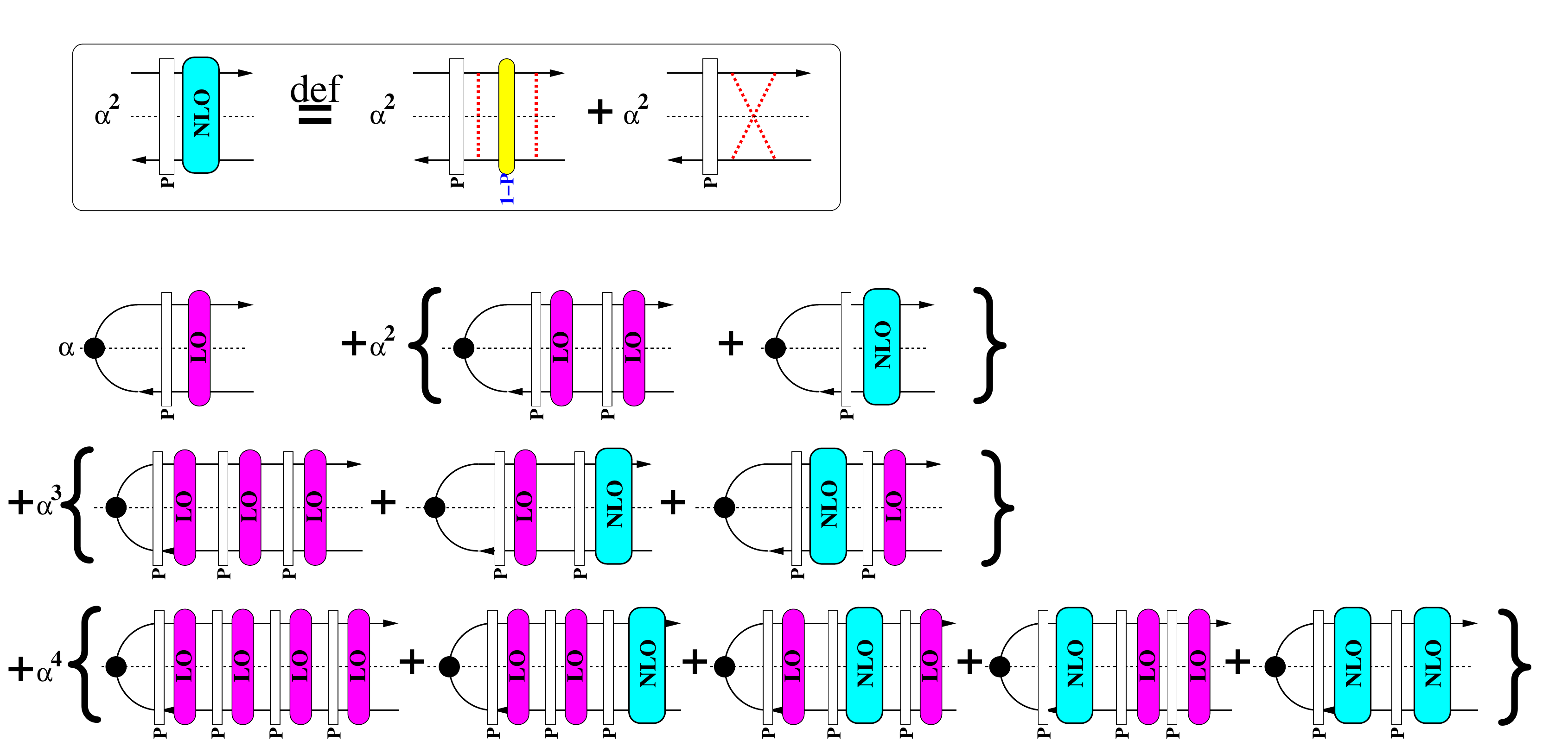}
\caption{
   General scheme of exclusive NLO insertions.
}
\label{fig:NLOins1}
\end{figure}

The very essence of the general methodology of the multiple ``insertion''
of the exclusive NLO kernel in the ladder of CFP scheme
is depicted in Fig.~\ref{fig:NLOins1}.
for the NLO MC distributions with up to four gluons.
The extension to more gluons is straightforward.
In Fig.~\ref{fig:NLOins1} terms with boxes marked as LO only,
form the traditional LO distributions being the product of the LO kernels,
times Sudakov formfactor resuming virtual corrections,
see below for more details.
Note that LO box includes trivial single real parton phase space.
On the other hand, the NLO box, which shows up for the first time for 2 real gluons
and is defined in the top part%
\footnote{Of course, we keep in mind that NLO kernels include
  virtual contribution within the single real parton phase space,
  similarly as LO box.}
of Fig.~\ref{fig:NLOins1}, features two real parton phase space.
The case of 2 gluons in Fig.~\ref{fig:NLOins1}
is trivial, because it is equivalent to the definition
of the NLO kernel.
The first nontrivial term in the decomposition of Fig.~\ref{fig:NLOins1}
is that for 3 real gluons,
where the two-gluon NLO part inserted either
before or after the LO kernel.
In case of four gluons in Fig.~\ref{fig:NLOins1}
we have 3 terms with one NLO insertion in 3 possible
ways and a double NLO insertion for the first time,
albeit without any LO spectators.
The extention of this pattern to any number of real gluons is not difficult.

The real problem is how to define,
or rather deduce from the original matrix element,
the convolution of many LO and NLO
factors within the standard Lorentz invariant phase space,
such that the infinite sum of such convolutions
obeys {\em exactly} the LO+NLO evolution equation of eq.~(\ref{eq:evoleq}).

At first glance the above task may seem unfeasible.
Nevertheless, we shall show in the following how to do it,
and demonstrate the first Monte Carlo implementation,
albeit only for the bremsstrahlung subset of the diagrams of the previous Sections.
Not surprisingly, the solution is found by examining carefully the original
$n$-parton matrix element coming directly from the Feynman diagrams,
and the procedures of the QCD factorization theorems used to extract
the ``idealized'' LO+NLO part from it.

\subsection{NLO insertion for DGLAP evolution with inclusive kernels}

We start the above difficult task with the warm-up exercise in which
the NLO insertions are defined within
the traditional iterative solution of the DGLAP evolution equation
with LO+NLO inclusive kernels.
The corresponding simple Markovian MC program
with the inclusive LO+NLO kernels will also serve
as a valuable testing tool of a more complicated new parton shower MC
with the exclusive LO and NLO kernels.

\begin{figure}[!ht]
\centering
   \includegraphics[width=120mm]{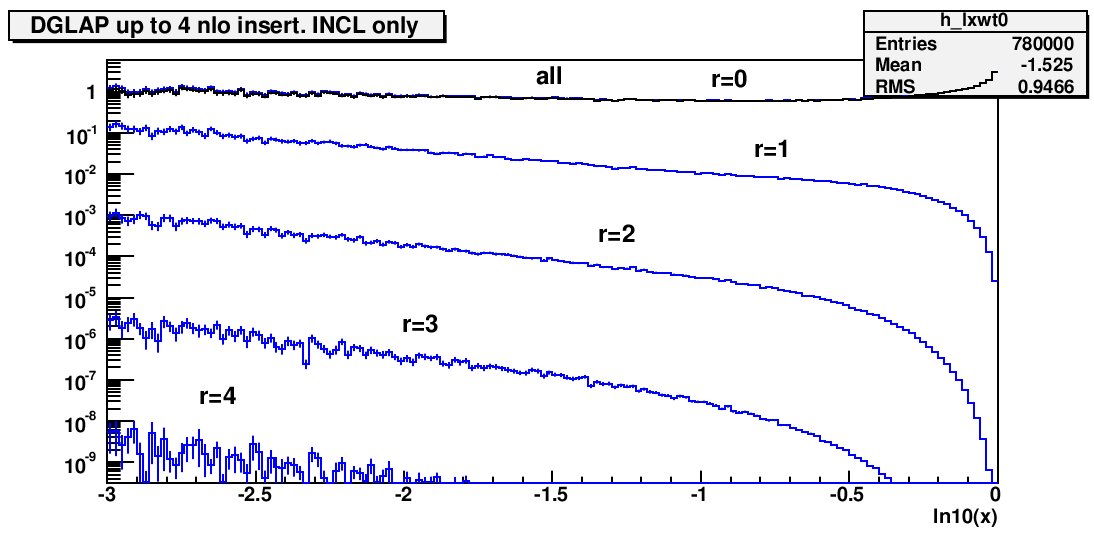}
\caption{
   Contributions to $D(t,x)$ for LO, LO+NLO
   and separate slices to NLO for $r=0,1,2,3,4$ NLO insertions
   from the Markovian MC with the {\em inclusive} kernels.
}
\label{fig:Slices}
\end{figure}

Changing slightly notation,
the traditional NLO DGLAP evolution reads
\begin{equation}
  \partial_t D(t,x) = \bar\alpha\; (\Peu \otimes D(t))(x)=
                      \bar\alpha  \int_0^1 dz dx\; \delta_{x=zy}\; \Peu(z) D(t,y),
\end{equation}
where the inclusive LO+NLO kernel is
$
  \Peu(z)= \Peu^{(0)}(z) + \bar\alpha \Peu^{(1)}(z), \bar\alpha=C_F\frac{\alpha}{\pi},
$
$t=\ln Q=\ln\mu$ and  $\alpha$ is non-running.
The NLO part of the kernel ${\bar\alpha}^2 \Peu^{(1)}(z)=\Peu^N(z)$
is exactly that of eq.~(\ref{eq:PN}).
The iterative solutions of the above equation reads as follows%
\footnote{Virtual LO corrections
  regularizing soft divergence $1/(1-z)$ are resummed into Sudakov
  formfactor $\exp(-S)$.
  The NLO part in $S$ is set to zero and $\int \Peu(z)dz \neq 0$. }
\begin{equation}
\label{eq:Dtx}
\begin{split}
&D(t,x) 
= e^{-S} \delta_{x=1}
 +e^{-S} \sum_{n=1}^\infty 
   \left( \prod_{i=1}^n \int_{t_{i-1}}^t \!\! dt_i\;  
     \big(\bar\alpha \Peu^{(0)}_\theta(z_i) +{\bar\alpha}^2 \Peu^{(1)}(z)\big) \right)
     \delta_{x=\prod z_j},
\\&
 \Peu^{(0)}(z)= \left(-\ln\frac{1}{\delta}+\frac{3}{4} \right) \delta_{x=1}
           +\Peu^{(0)}_\theta(z),\quad
 \Peu^{(0)}_\theta(z) =  \frac{1+z^2}{2(1-z)} \theta_{1-z>\delta},
\\&
S= \bar\alpha (t-t_0) \left(\ln\frac{1}{\delta}-\frac{3}{4} \right).
\end{split}
\end{equation}
In the Markovian MC exercise parton multiplicity
$n$ and the entire MC event $\{t_i,z_i, i=1,2,...,n\}$ is generated,
using only LO kernel $\Peu^{(0)}(z)$.
The NLO content is introduced using MC weight
\begin{equation}
 w=\frac{\prod_{i=1}^n (\Peu^{(0)}_\theta(z_i)+\alpha\Peu^{(1)}(z_i))}%
          {\prod_{i=1}^n \Peu^{(0)}_\theta(z_i)}
  = 
     \frac{\sum_{ \{ d \} } \alpha^{r(d)} \prod_{i=1}^n\Peu^{(d_i)}(z_i)}%
          {\prod_{i=1}^n\Peu^{(0)}_\theta(z_i)},
\end{equation}
where $\{ d\}$ is the set of all $2^n$
partitions $d=(d_1,d_2,d_3,...,d_n)$, $d_i=0,1$.
$r(d)=\sum_i d_i$ is the number of NLO factors
in a given partition.
The decomposition in Fig.~\ref{fig:NLOins1} is essentially that of
the numerator in the above expression truncated to
terms of the order $\alpha^n$, $n\leq 4$.

Fig.~\ref{fig:Slices} presents result from the MC run for
evolution from $Q_0=e^{t_0}=1$GeV to $Q=e^t=1000$GeV and $\alpha=0.2$.
In the plot we see LO result for $r=0$, the NLO result, 
and the slices with the defined number of the NLO insertions
being $r=0,1,2,3,4$.

The important conclusion from this introductory exercise
is that for any practical purpose we may truncate the sum%
\footnote{For the subset of the bremsstrahlung diagrams under discussion.}
over NLO insertions to $r(d)\leq 3$ or even $r(d)\leq 2$.

\subsection{Introductory step: extraction of LO from the exact matrix element}

\begin{figure}[!ht]
\centering
   \includegraphics[width=125mm]{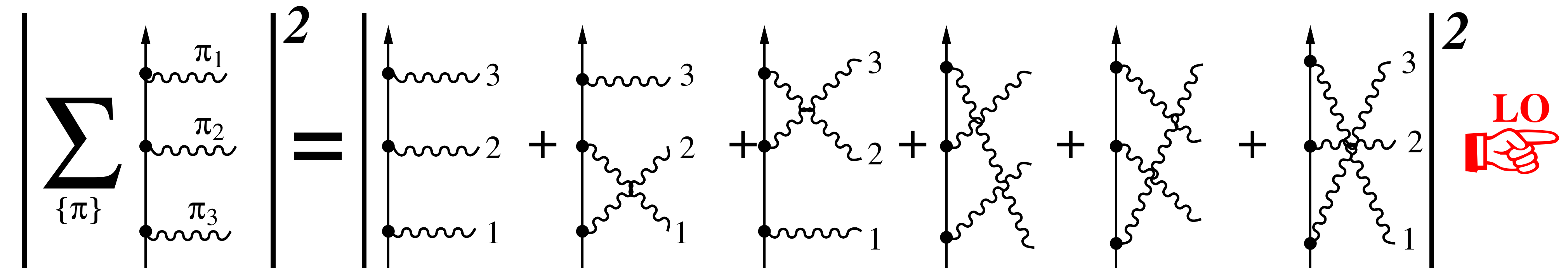}
   \includegraphics[width=125mm]{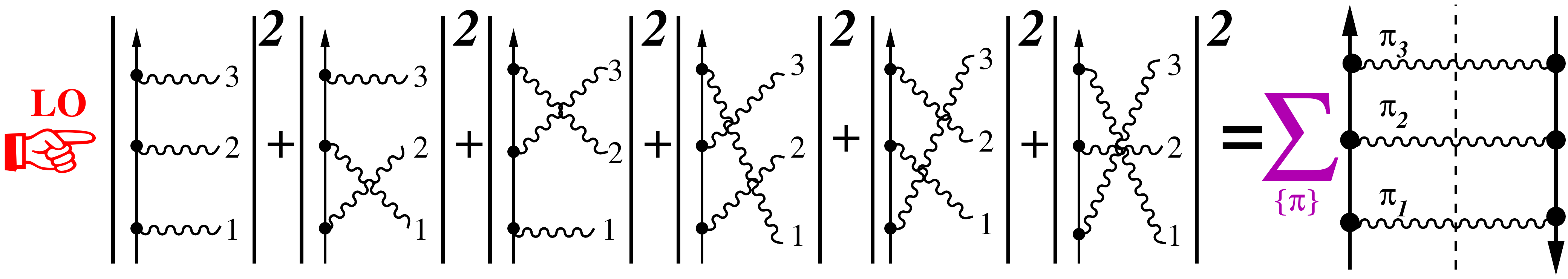}
   \includegraphics[width=40mm]{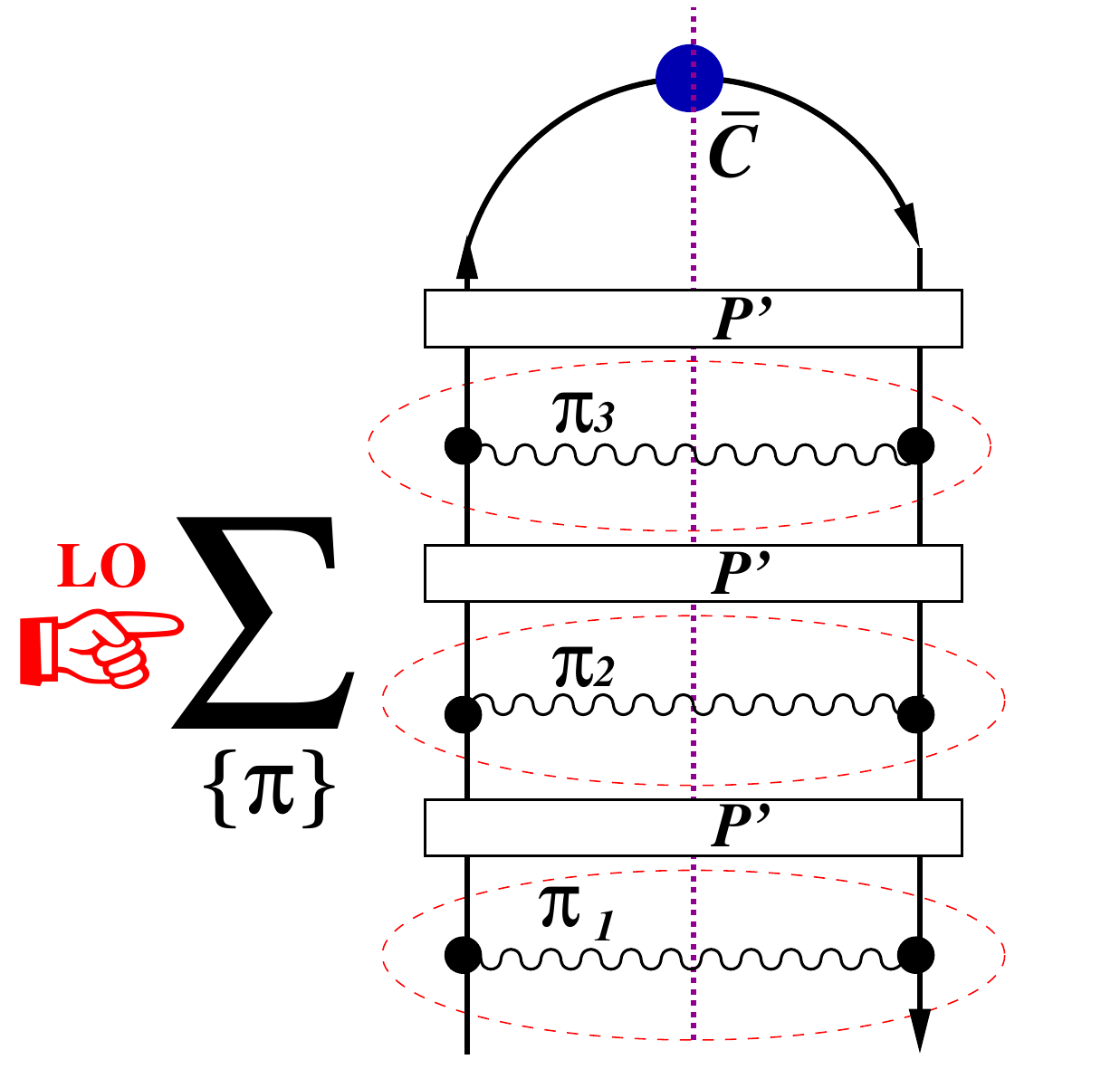}
\caption{
   Extraction of LO from the exact matrix element for 3 real gluons.
}
\label{fig:extractLO3}
\end{figure}

As already stressed, the main problem in devising LO+NLO exclusive
distribution is how to combine NLO insertions and LO parts.
We have solved it by examining once again how LO+NLO distributions are extracted
from the exact multiparton matrix element (i.e. from Feynman diagrams).
This cannot be done without careful re-examination of this procedure for
$n$ partons at the seemingly trivial LO level.

Let us do it for $n=3$, that is for the emission of
3 gluons out of an emitter quark.
The whole procedure is depicted in Fig.~\ref{fig:extractLO3}
and it goes in two steps:
we start with $n!=6$ diagrams squared defined in the entire phase space
(and $1/n!$ Bose factor in front).
The explicit squared sum over $n!$ permutations
$\{ \pi \}=(123),(213),(132),(231),(312),(321)$
is expanded into $(n!)^2=36$ terms.

In the axial gauge in the LO approximation it is allowed
to drop all interference terms~\cite{lipatov:75},
so in the {\em 1st step} we do it retaining out of $(n!)^2$ terms only
$n!=6$ diagonal terms depicted in the second line in
Fig.~\ref{fig:extractLO3}.
Each of the $n!$ diagrams covers approximately one simplex
$v>v_{\pi_3}>v_{\pi_2}>v_{\pi_1}>v_0$,
where $v_i=k^T_i/\sqrt{\alpha_i}=k^-_i$.
This for example is illustrated in the upper left plot in
Fig.~\ref{fig:Br2S}.
However, as clearly seen in this figure, the shape of the distribution
is {\em fuzzy} at the border of the simplices $v_1\sim v_2$.
In other words the bremsstrahlung matrix element (squared)
of the single Feynman diagram in the axial gauge
features what we call {\em fuzzy ordering}.

In the {\em 2nd step}
the insertion of the projection operator $P$ replaces the fuzzy
ordering (at the simplice boundaries) by the sharp one.
In fact the $\overline{MS}$ scheme dictates the LO distribution
to be sharp along the lines of equal transverse momentum%
\footnote{
  In the soft limit $\alpha_i\to 0$
  the exact matrix element,
  summed over $n!$ diagrams,
  is not able to tell us whether the LO distribution
  (also summed up over $n!$ diagrams)
  with sharp ordering using $k^T_i$ or $v_i$ variables is better,
  as they all coincide in this limit.},
$Q>k^T_{\pi_3}>k^T_{\pi_2}>k^T_{\pi_1}>Q_0$,
as also seen in Fig.~\ref{fig:Br2S} for $n=2$.

The important lesson from the above LO considerations is
that it is wise to keep the complete phase space,
without any artificial division into simplices suggested by the
so called ordering property of LO-approximated
matrix element in the variables like $v_i$ or $k^T_i$,
but instead to associate such a division
{\em only} with the properties of the particular diagram.
At the LO level there is, of course, one-to-one correspondence among
$n!$ squared diagrams and $n!$ simplices in $k^T$ in the phase space.
This may easily be the source of the confusion,
because this one-to-one correspondence exists only at the LO level
and breaks down beyond the LO approximation,
when one gradually goes back towards exact matrix element,
for instance by recovering terms contributing at the NLO level,
or replacing sharp ordering of LO
by the fuzzy ordering of the exact matrix element.
This we have already seen at work when studying properties
of the NLO unintegrated kernel in the previous section.

\begin{figure}[!ht]
\centering
   \includegraphics[width=125mm]{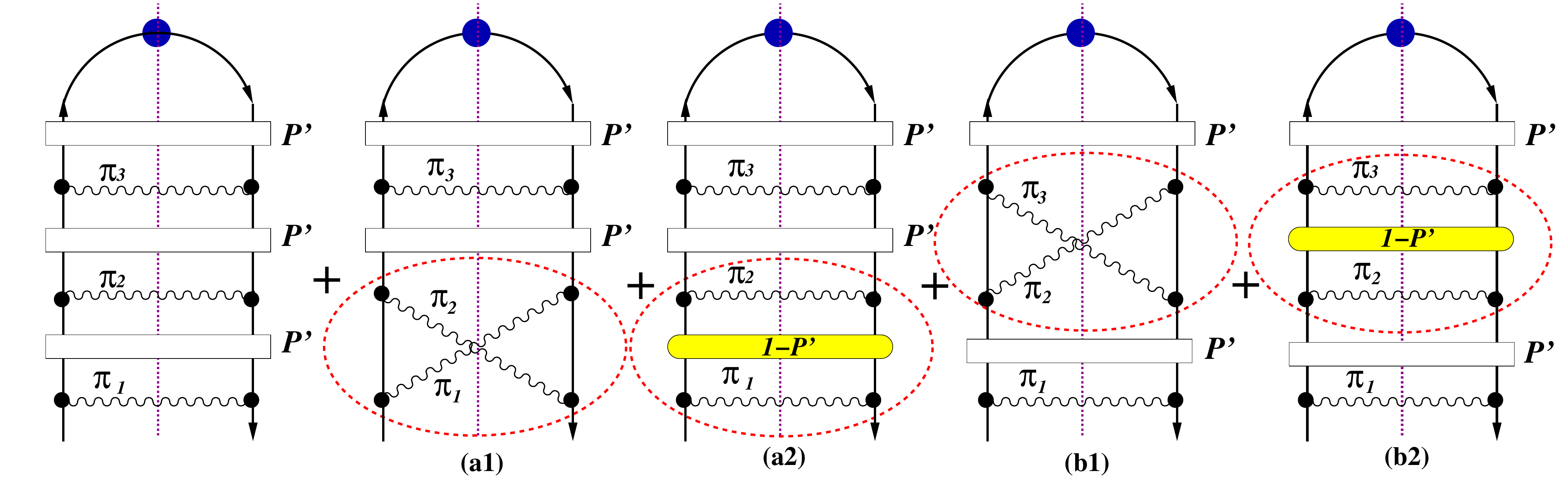}
\caption{
   Additional NLO contributions forming single NLO insertion
   (dashed ellipsis) for $n=3$ gluon emissions.
}
\label{fig:ladder3ins1}
\end{figure}

\subsection{NLO insertion and LO spectator for $n=3$}
The necessity of combining LO kernel with NLO insertion is encountered
for the first time for 3 gluons.
Let us analyze the problem and find solution in 
this simplest beyond-the-LO case.

Following similar factorization procedure for 3 gluons
as in the LO discussion of the previous section,
let us now retain the contributions of the NLO class
in addition to the product of LO kernels.
The new NLO objects are of two kinds:
(i) the interference contributions among
two adjacent emissions in the Feynman
diagram, which were neglected in step 1
of the previous LO example,
see (a1) and (b1) in Fig.~\ref{fig:ladder3ins1},
and
(ii) the contribution due to
restoring of the ``fuzzy ordering'' of the exact matrix element
(removing $\Pbbm$ operator responsible for the sharp ordering)
for the same two adjacent emissions,
see (a2) and (b2) in Fig.~\ref{fig:ladder3ins1}.

The extra terms of two kinds show up for 3 gluons 
in two possible locations in the emission ladder,
as shown in Fig.~\ref{fig:ladder3ins1},
and form in a natural way a single NLO insertion in the presence
of a single LO spectator,
(a1)+(a2) and (b1)+(b2) in Fig.~\ref{fig:ladder3ins1}.
As seen in Fig.~\ref{fig:ladder3ins1} the LO spectator can be positioned
either before or after NLO insertion.
Let us now translate the above description into formulas.

First of all, we upgrade slightly our LO MC model, by means of adding
into the game the azimuthal angle of the emitted partons
and restoring four momenta of all partons.
This is done with the help of the usual mapping of the evolution variables
into four momenta:
$ k_j^T=e^{t_i},\; k_j^+=2(x_{j-1}-x_j)E_h,\; k_j^-=(k_j^T)^2/k_j^+$.
The distribution of the LO MC for $n=3$ emissions takes now
the following shape within the standard phase space, cf. eq.~(\ref{eq:Dtx}),
\begin{equation}
\begin{split}
&D^{L}_3(t,x) 
=e^{-S}
\int \frac{d^3 k_3}{2 k^0_3}
\int \frac{d^3 k_2}{2 k^0_2}
\int \frac{d^3 k_1}{2 k^0_1}\;
   \delta_{1-x=\alpha_1+\alpha_2+\alpha_3}\; 
   \prod_{i=1}^3 \theta_{e^t>|\bk_i|>e^{t_0}}
\\&~~~~~~~~~~
  \rho^L(k_3|x_2)\;  \rho^L(k_2|x_1)\;  \rho^L(k_1|x_0)\;
  \theta_{|\bk_3|>|\bk_2|>|\bk_1|},
\\&
\rho^L(k_i|x_{i-1}) = \frac{\bar\alpha}{16\pi} \frac{1+z_i^2}{(1-z_i)|\bk_i|^2}.
\end{split}
\end{equation}
Remembering lessons from the discussion of the LO factorization
we rewrite the above with the explicit sum over Feynman diagrams squared%
\footnote{With sharp ordering and no interferences.}
\begin{equation}
\begin{split}
&D^{L}_3(t,x) 
=  e^{-S}
\int \frac{d^3 k_3}{2 k^0_3}
\int \frac{d^3 k_2}{2 k^0_2}
\int \frac{d^3 k_1}{2 k^0_1}\;
   \delta_{1-x=\alpha_1+\alpha_2+\alpha_3}\; 
   \prod_{i=1}^3  \theta_{e^t>|\bk_i|>e^{t_0}}
\\&~~~~~~~
  \frac{1}{3!} \sum_{\{\pi\} }\;
  \rho^L(k_{\pi_3}|x^\pi_{2})\;
  \rho^L(k_{\pi_2}|x^\pi_{1})\;
  \rho^L(k_{\pi_1}|x_0)\;
   \theta_{|\bk_{\pi_3}|>|\bk_{\pi_2}|>|\bk_{\pi_1}|},
\end{split}
\end{equation}
where $k^\mu_i=(k^0_i,\bk_i,k^3_i)$, $\alpha_i=(k^0_i+k^3_i)/(2E_h)$,
$x^\pi_1=x_0-\alpha_{\pi_1},
x^\pi_2=x_0-\alpha_{\pi_1}-\alpha_{\pi_2}$.
Note that inclusion of the lighcone variable $x_{i-1}$ of
the quark emitter (prior to emission)
into $\rho^L(k_{\pi_i}|x^\pi_{i-1})$ was necessary
in order to define $z_i=x_i/x_{i-1}=1-\alpha_i/x_{i-1}$
in terms of the four momenta $k_i^\mu$.

We are now able to write down the LO+NLO contribution:
\begin{equation}
\label{eq:DLN}
\begin{split}
&D^{L+N}_3(t,x)
=\frac{1}{3!} e^{-S}
\int \frac{d^3 k_3}{2 k^0_3}
\int \frac{d^3 k_2}{2 k^0_2}
\int \frac{d^3 k_1}{2 k^0_1}\;
\\&~~~~~~~~~~~~~~~~\times
  \delta_{x_0-x=\alpha_1+\alpha_2+\alpha_3}
  \prod_i \theta_{e^t>|\bk_i|>e^{t_0}}\;
  \rho^{L+N}_3(k_3,k_2,k_1),
\\&
\rho^{L+N}_3(k_3,k_2,k_1)=
\\&~~~~~
\sum_\pi\left(
   \rho^{L}_3(k_{\pi_3},k_{\pi_2},k_{\pi_1})
  +\rho^{N}_{3a}(k_{\pi_3},k_{\pi_2},k_{\pi_1})
  +\rho^{N}_{3b}(k_{\pi_3},k_{\pi_2},k_{\pi_1}) \right),
\end{split}
\end{equation}
where
\begin{equation}
\begin{split}
&\rho^{L}_3(k_3,k_2,k_1)=
  \rho^L(k_3|x_2)\;  \rho^L(k_2|x_1)\;  \rho^L(k_1|x_0)\;
   \theta_{|\bk_3|>|\bk_2|>|\bk_1|},
\\&
 \rho^{N}_{3a}(k_3,k_2,k_1)
  =\rho^L(k_3|x_2)\;  \rho^{N}(k_2,k_1|x_0)\;
   \theta_{|\bk_3|>\max(|\bk_2|,|\bk_1|)},
\\&
 \rho^{N}_{3b}(k_3,k_2,k_1)
  =\rho^{N}(k_3,k_2|x_1)\;\rho^L(k_1|x_0)\;
   \theta_{\max(|\bk_3|,|\bk_2)>|\bk_1|}.
\end{split}
\end{equation}
By examining eq.~(\ref{eq:b2N}) we identify
\begin{equation}
\rho^N(k_{2},k_{1}|x')=
  b^N_2(\bk_2,\alpha_2/x', \bk_1,\alpha_1/x')\; 2\theta_{|\bk_2|>|\bk_1|},
\end{equation}
where $x'$ of the emitter rescales $\alpha$'s,
but transverse momenta $\bk_i$ are unchanged%
\footnote{Factor $ 2\theta_{|\bk_2|>|\bk_1|}$ could be omitted
  -- its sole role is to reduce slightly combinatorics.}.

Two important points about the above equations:
(i) For $\rho^{N}_{3b}$, where LO spectator $k_1^\mu$ proceeds NLO insertion,
the ordering $|\bk_2|>|\bk_1|$ is violated,
i.e. $\rho^{N}_{3b}$ contributes to two
out of six simplices in the $|\bk_i|$ space%
\footnote{This is contrary to $\rho^{L}_3$ and $\rho^{N}_{3a}$,
  which contribute to one simplex in $t_i$ variables --
  in other words $\rho^{N}_{3b}$ corrects for the ``sharp ordering''
  of LO in the $|\bk_2|\simeq|\bk_1|$ region.}.
This is of course general phenomenon and in the case of more such LO spectators
$k_j^\mu$ of the NLO insertion may swap 
with any one of these proceeding LO spectators.
This shows that the sum $\sum_\pi$ is important and instrumental.
(ii) The validity of the exact NLO DGLAP%
\footnote{Limited, of course, to our subset of diagrams.}
evolution of eq.~(\ref{eq:evoleq}) requires that
\begin{equation}
\label{eq:LN3}
\begin{split}
&D^{L+N}_3(t,x)=
   e^{-S} {\bar\alpha}^3 (t-t_0)^2
\\&~~~~~~~~~~~~
  \times\left[
   \frac{1}{3!} \Peu^L_\theta \otimes  \Peu^L_\theta \otimes  \Peu^L_\theta(x)
  +\frac{1}{2!} \Peu^L_\theta \otimes  \Peu^N (x)
  +\frac{1}{2!} \Peu^N \otimes  \Peu^L_\theta(x)
   \right]
\end{split}
\end{equation}
is true.
We have checked by analytical and numerical integration
that the above is {\em almost} valid. The only problem is
with integration limits for the NLO subintegral.
Even for simpler $\rho^{N}_{3a}$ the integration limit
of $|\bk_1|$ (evolution starting point) is $|\bk_1|>e^{t_0}$,
while in eq.~(\ref{eq:PN}) this limit is exactly zero.
One method to overcome this problem and to assure
the exact validity of eq.~(\ref{eq:evoleq}) is
to introduce another substantially lower starting point
$t_M$ of the evolution; in the interval $(t_0,t_M)$
evolution is performed with LO kernels and NLO insertions are activated
only within $(t_0,t)$. The error in eq.~(\ref{eq:LN3})
will be reduced to a mere $\sim e^{t_M-t_0}\ll 1$
(i.e. power suppressed).

\subsection{Efficient evaluation of the Monte Carlo weight}
The structure of the MC weight for NLO distribution
for $n$ partons,
\begin{equation}
 w_n = \frac{\rho^{L+N}_n}{\rho^{L}_n}=
\frac{\sum\limits_{\{\pi\}} \bigg(
  \rho^{L}_n(k_{\pi_1},\dots,k_{\pi_n})
 +\sum\limits_{r=1}^{n/2}\sum\limits_{\{\omega\}}
  \rho^{N}_{NR\omega}(k_{\pi_1},\dots,k_{\pi_n})
             \bigg)}%
      {\sum_{\{\pi\}}  \rho^{L}_n(k_{\pi_1},\dots,k_{\pi_n})},
\end{equation}
includes sum over number of NLO insertions $r$
and sum over set of $\{\omega\}$ numbering
all possible ways of placing $r$ insertions within the $n$-rung ladder.
For many partons
it would take prohibitively long CPU time to evaluate the above sums
(especially $\sum_{\{\pi\}}$) for every MC event.
However, as already pointed out we may cut at $r\leq 2$.
Moreover, for a given MC point $(k_{\pi_1},\dots,k_{\pi_n})$
only one special permutation $\pi_k$ with the right ordering contributes
in the LO part  $\sum_{\{\pi\}} \rho^{L}_n$.
In the NLO part there is a lot of zero contributions as well.
We exploit this fact and reorder permutations with respect to $\pi_k$
using group product relation $\pi=\pi^k\pi'$
\begin{equation}
 w_n = 1
 +\frac{1}{\rho^{L}_n(k_{\pi^k_1},\dots,k_{\pi^k_n})}
 \sum\limits_{r=1}^{n/2}\sum\limits_{\{\omega\}}
  \sum_{\{\pi'\}}
 \rho^{N}_{NR\omega}(k_{(\pi^k\pi')_1},\dots,k_{(\pi^k\pi')_n})
\end{equation}
and change summation order.
Next we restrict summation over
the set of all permutations $\{\pi'\}$ to a much smaller subset
$\{\pi^\omega\}$ of the permutations for which $\rho^{N}_{NR\omega}\neq 0$:
\begin{equation}
 w_n = 1
 +\frac{1}{\rho^{L}_n(k_{\pi^k_1},\dots,k_{\pi^k_n})}
  \sum\limits_{r=1}^{n/2}\sum\limits_{\{\omega\}}
  \sum_{\{\pi^\omega\}}
  \rho^{N}_{NR\omega}(k_{(\pi^k\pi^\omega)_1},\dots,k_{(\pi^k\pi^\omega)_n})
\end{equation}

How to organize $\sum_{\{\pi^\omega\}}$ such
that zero contributions are avoided?
In the general case this is a non-trivial task.
Details of the general solution will be published elsewhere.
Here let us only show it for $n=3$.
In this case in eq.~(\ref{eq:DLN}) there are two insertions
$\{ \omega \}= \{a,b\}$,
see (a1-a2) and (b1-b2) in Fig.\ref{fig:ladder3ins1};
for $a$ only one permutation enters
$\pi^a=(1,2,3)$, while for $b$
only two (out of six) permutations $\pi^b=(1,2,3),(2,1,3)$ 
do contribute:
\begin{equation}
\begin{split}
&w_3= 1+w^{N}_{3a}+w^{N}_{3b},
\qquad
w^{N}_{3a}
=\frac{
   \rho^{N}(\tilde{k}_2,\tilde{k}_1|x_0)
  }{
   \rho^L(\tilde{k}_{2}|x_1)\;
   \rho^L(\tilde{k}_{1}|x_0)}
  \theta_{\tilde{t}_2>t_{M}},
\\
&w^{N}_{3b}
=\frac{
   \rho^{N}(\tilde{k}_3,\tilde{k}_2|x_1)
  }{
   \rho^L(\tilde{k}_{3}|x_2)\;
   \rho^L(\tilde{k}_{2}|x_1)}
   \theta_{\tilde{t}_3>t_{M}}
+\frac{
   \rho^{N}(\tilde{k}_3,\tilde{k}_1|x_1^{\pi^b_2})
  }{
   \rho^L(\tilde{k}_{3}|x_2)\;
   \rho^L(\tilde{k}_{1}|x_0)}\;
  \frac{\rho^L(\tilde{k}_2|x_0)}{\rho^L(\tilde{k}_{2}|x_1)}
   \theta_{\tilde{t}_3>t_{M}},
\end{split}
\end{equation}
where $x_1^{\pi^b_2}=x_0-\alpha_2$ and
shorthand notation 
$ \tilde{k}_{i}\equiv k_{\pi^k_i},\; \tilde{t}_{i}\equiv t_{\pi^k_i}$ was used.

\begin{figure}[!ht]
\centering
  \includegraphics[width=80mm,height=130mm,angle=90]{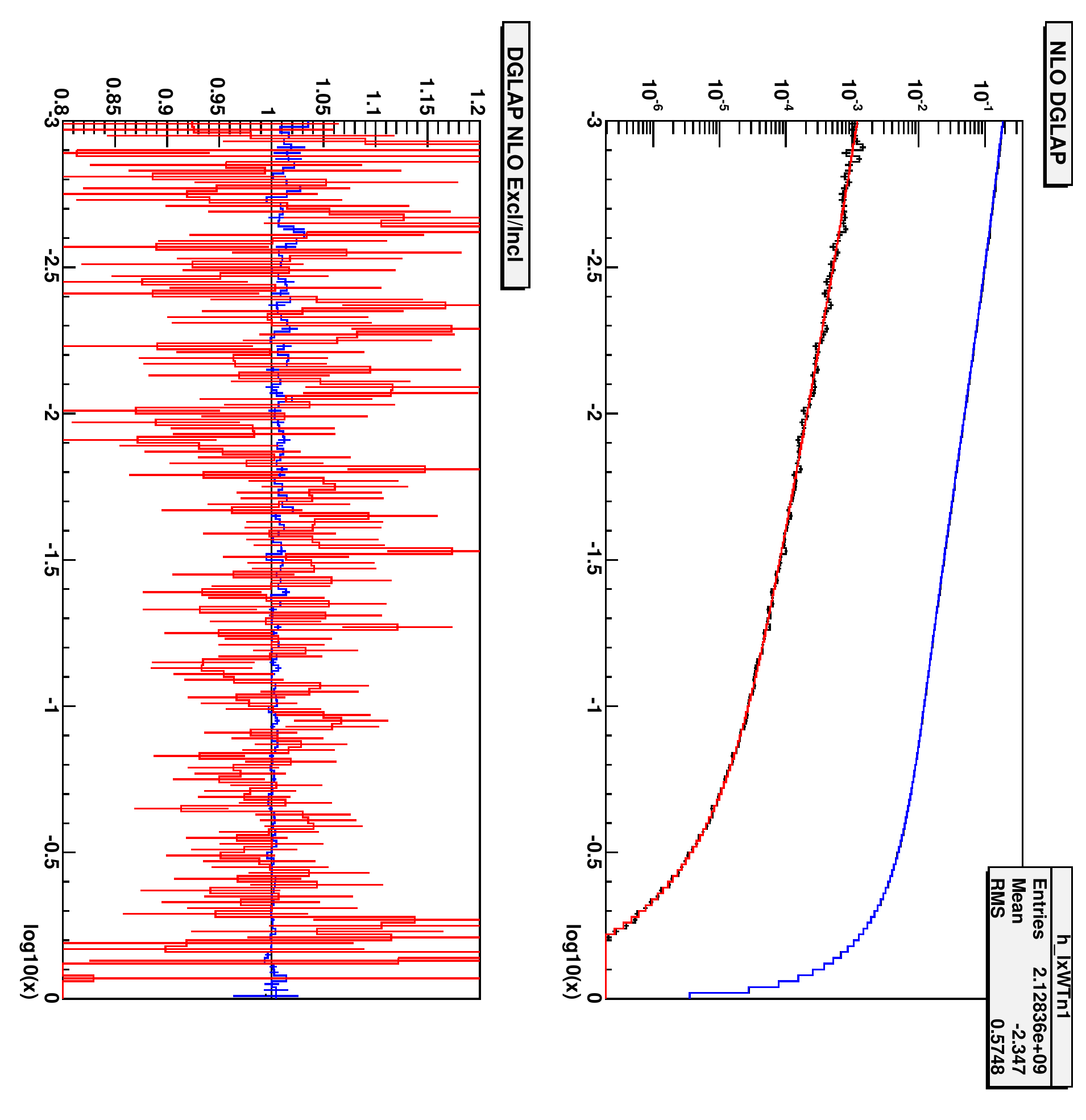}
 \caption{
   Comparison of MC results
   for NLO DGLAP evolutions modelled in the exclusive and inclusive ways.
   Contributions with $r=1,2$ of NLO insertions shown separately.
   Bremsstrahlung diagrams only. The  $\varepsilon$-term $T_2'$ is omitted.
   Total $2\cdot 10^{9}$ of MC events.
}
\label{fig:main}
\end{figure}

\begin{figure}[!ht]
\centering
   \includegraphics[width=80mm,height=130mm,angle=90]{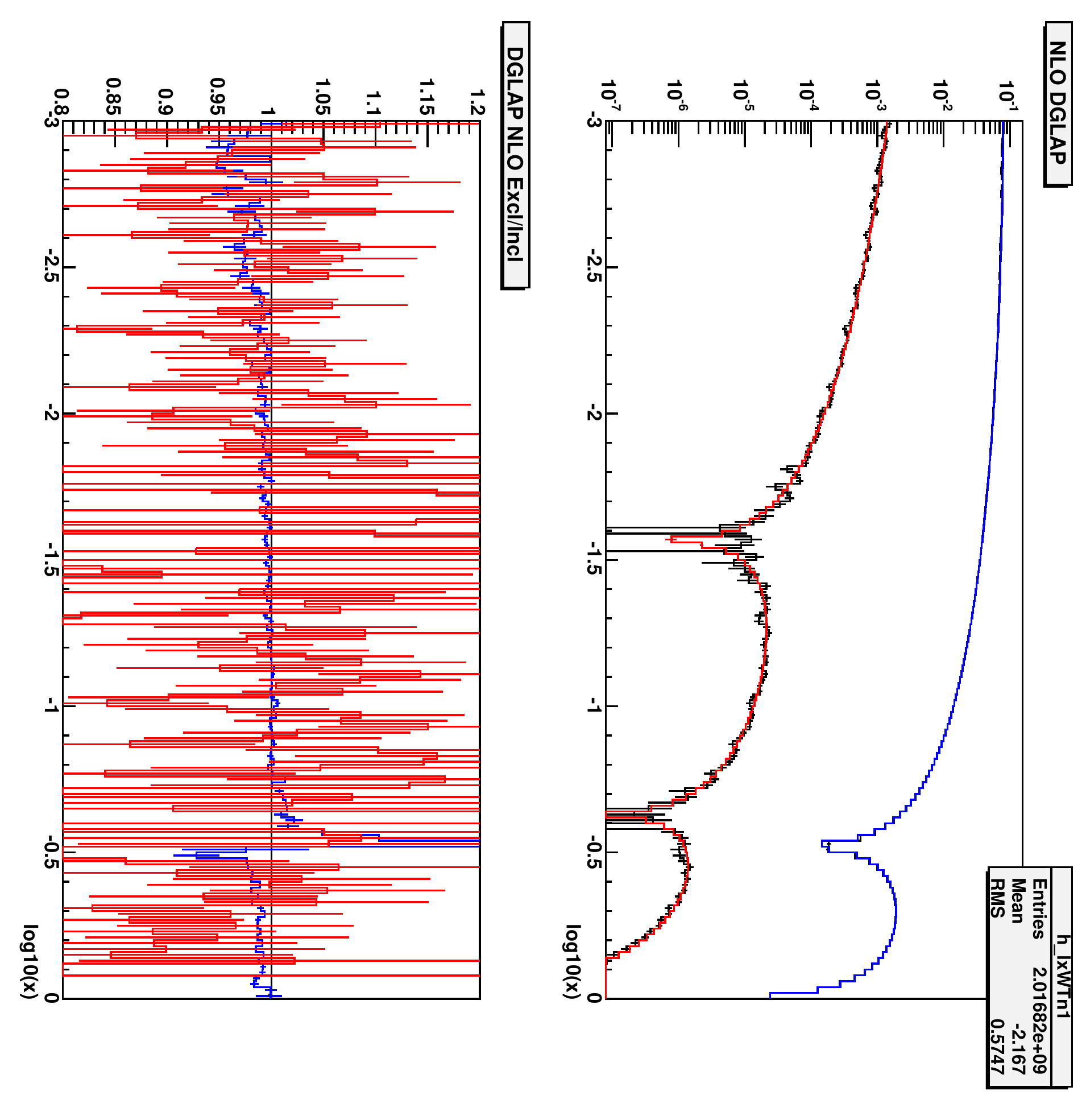}
\caption{
  Similar test as in Fig.~\protect\ref{fig:main}.
  This time $\varepsilon$ terms are included. $2\cdot 10^9$ events.
  Absolute value is plotted due to the use of the vertical logarithmic scale.
}
\label{fig:epsy}
\end{figure}

\begin{figure}[!ht]
\centering
   \includegraphics[width=39mm,height=110mm,angle=90]{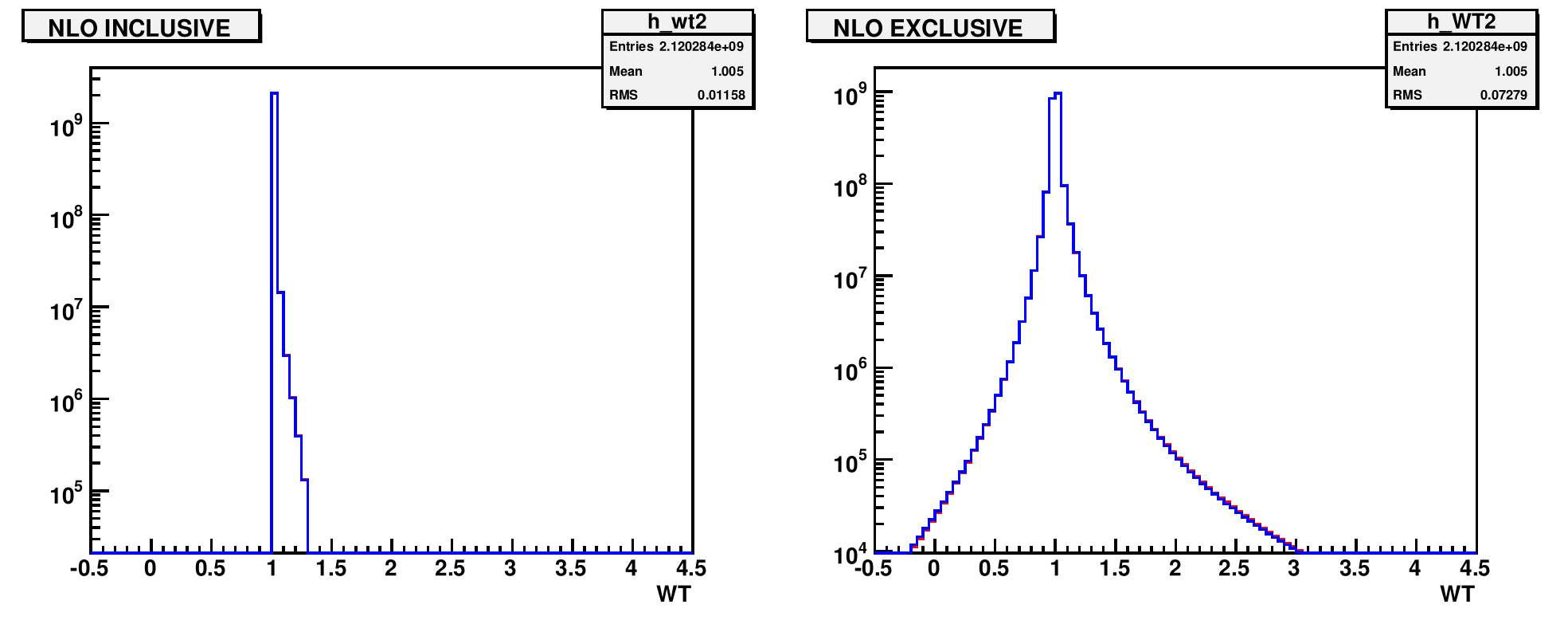}
   \includegraphics[width=37mm,height=80mm,angle=90]{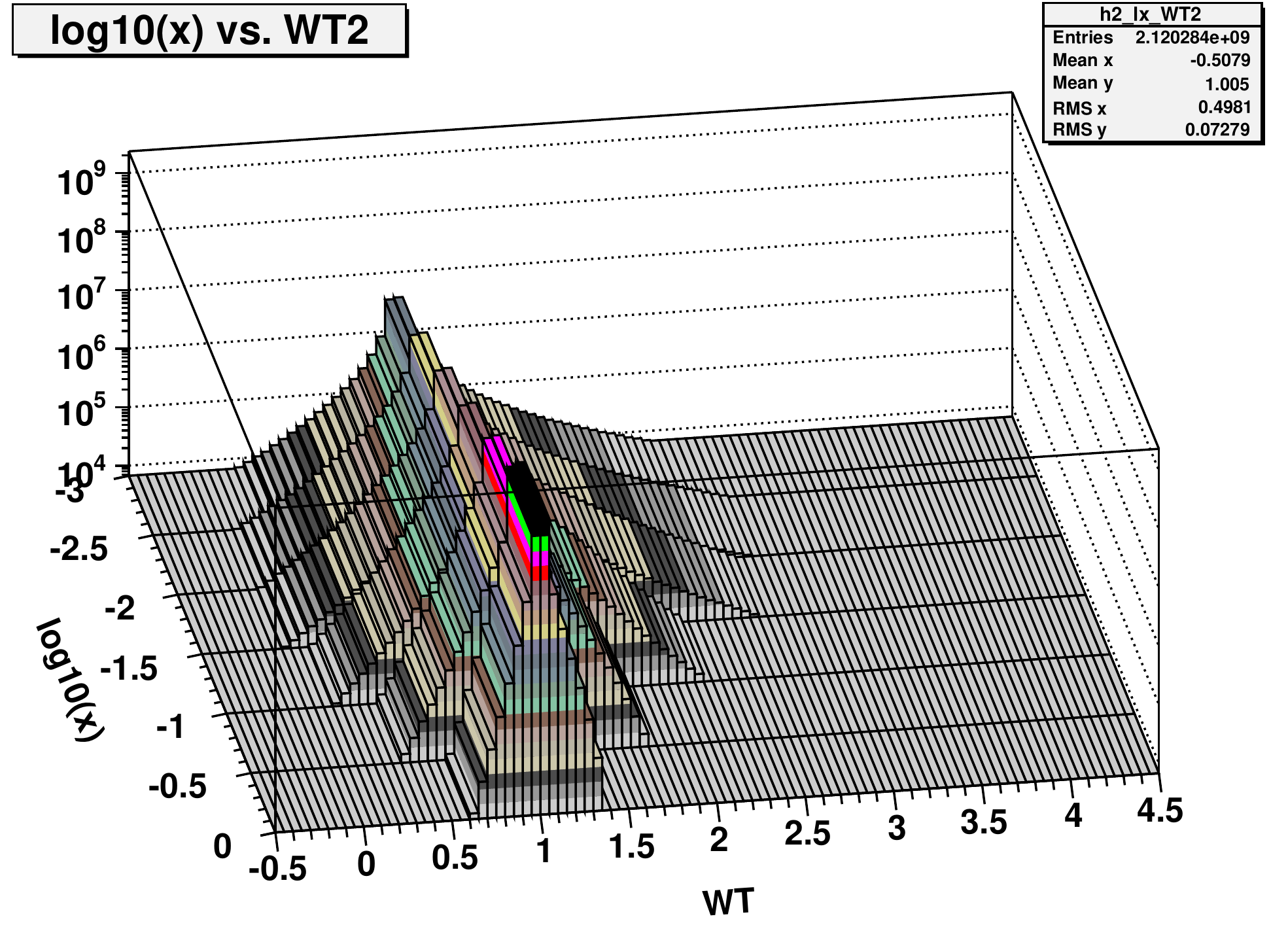}
\caption{ Weight distributions from LO+NLO Monte Carlo
  with inclusive kernels (upper left) and exclusive kernels
  (upper right). 
  The $x$ dependence of the weight distribution in the latter
  case is also shown in the lower 2-dim. plot.
}
\label{fig:wtdist}
\end{figure}

\subsection{Main numerical result}

In the previous section we have sketched the structure
of the LO+NLO distribution for the MC with exclusive kernels.
The above scheme of NLO insertions and the methodology
of calculating MC weight was worked out and tested for
up to two NLO insertions for any number of partons
in the case of the bremsstrahlung diagrams.
In Fig.~\ref{fig:main} we demonstrate
with the high precision numerical tests that
the above LO+NLO Monte Carlo scheme works well in practice.
The agreement of two MC results is within statistical MC error%
\footnote{%
   Obviously, the statistical MC errors are exaggerated
   in this and the following plots by factor 3 or more due to dense
   binning in $\log_{10}(x)$.}
which for $r=1$ is $\sim 10^{-3}$ and for $r=2$ is $\sim 10^{-4}$,
relative to total LO+NLO.
This is the most important result of this report.

More on what is in the plot in Fig.~\ref{fig:main}:
(i)
Both evolutions are implemented using the same underlying Markovian LO MC%
\footnote{It can be put easily on top of
non-Markovian constrained MC~\cite{Jadach:2005bf}.}.
(ii)
Terms due to $\varepsilon$ part of $\gamma$-traces are omitted.
The same exercise,
with roughly the same precision tag as in Fig.~\ref{fig:main},
but with $\varepsilon$ terms included
is shown in Fig.~\ref{fig:epsy}.
(iii)
MC weights are positive, weight distributions are very reasonable,
see weight distributions in Fig.~\ref{fig:wtdist}
(iv)
Evolution ranges from 10GeV to 1TeV;
LO Prue-evolution, starting from $\delta(1-x)$ at 1GeV and ranging to 10GeV,
provides initial $x$-distribution for the LO+NLO continuation.
(v)
As before, only $C_F^2$ part of gluonstrahlung and
non-ruining $\alpha_S$ are used.
(vi)
NLO virtual corrections are omitted.
However, the methodology of including them is at hand, it was
already used to include $\varepsilon$ terms (Fig.~\ref{fig:epsy}).

\section{Conclusions and outlook}

The NLO evolution
of the parton distribution functions in QCD is a fundamental tool
in the lepton-hadron and hadron-hadron collider data analysis.
It has been formulated for the inclusive (integrated)
PDFs with the help of inclusive NLO kernels.
We report here on the ongoing work
in which NLO DGLAP evolution is performed for 
the exclusive (fully unintegrated)  multiparton distributions, ePDF's,
with the help of the exclusive kernels.
These kernels are calculated within the two-parton phase space using
Curci-Furmanski-Petrionzio factorization scheme for bremsstrahlung
subset of the Feynman diagrams of the non-singlet evolution kernel.
The multiparton NLO-compatible distribution (with multiple use of
the exclusive NLO kernels) is implemented in the Monte Carlo program
simulating multi-gluon emission from a single quark emitter.
High precision MC results show that
the new scheme works perfectly well in practice and is fully compatible
(equivalent at the inclusive level) with the traditional inclusive
NLO DGLAP evolution.
Once completed, this Monte Carlo module is aimed as a building
block for the NLO parton shower Monte Carlo, for $W/Z$ production
at LHC and for $ep$ scattering, and as a starting point
for other perturbative QCD based Monte Carlo projects.
Presented study is limited to $C_F^2$ part of bremsstrahlung diagrams.
In the immediate future the remaining non-single diagrams will be included
and the singlet case will be also worked out.

\vspace{4mm}
\noindent
{\bf Acknowledgments}\\
One of the authors (S.J.) would like to acknowledge warm hospitality
and support of
Institute of Nuclear Physics, NCSR Demokritos, Athens,
and
Max-Planck-Institut f\"{u}r Physik, M\"{u}nchen,
where part of the presented work was done.
Both of us wish to thank CERN Theory Unit for hospitality and support.


\providecommand{\href}[2]{#2}

\end{document}